\documentclass[a4paper,11pt]{article}
\pdfoutput=1

\usepackage{jheppub}

\usepackage{amsmath,amsthm,hyperref}
\usepackage{bbold}
\usepackage{cleveref}
\usepackage{hyperref}
\usepackage{subcaption}
\usepackage{array}
\usepackage{wasysym}

\usepackage{tikz}
\usetikzlibrary{decorations.pathmorphing}
\tikzset{grav/.style={decorate, decoration=snake}}
\tikzset{cross/.style={path picture={ 
  \draw[black]
(path picture bounding box.south east) -- (path picture bounding box.north west) (path picture bounding box.south west) -- (path picture bounding box.north east);
}}}

\newcolumntype{V}{>{\centering\arraybackslash} m{.4\linewidth} }

\newcommand{\RR}{\mathbb{R}} %Reals
\DeclareMathOperator{\sech}{sech}

\newcommand{\piD}[1]{\mathcal{D}#1} % Path integral measure
\title{A view of the bulk from the worldline}
\author{Henry Maxfield}
\affiliation{Physics Department, McGill University	\\	Montr\'eal, QC H3A 2T8, Canada}
\emailAdd{henry.maxfield@physics.mcgill.ca}

\abstract{
A new method to compute correlation functions in AdS$_{d+1}$ in general dimension is introduced, considering a particle quantised in the worldline formalism of quantum field theory, coupled to bulk fields, in particular gravity, quantised in the standard manner. This gives a systematic two-parameter perturbative expansion organised by small parameters $G_N$ and the inverse mass of the particle, complementary to the usual Witten diagram expansion. This connects closely to CFT language, with the geodesic Witten diagram representation of global conformal blocks emerging naturally, and for two dimensions gives a bulk representation of the Virasoro block, with a systematic method for computing quantum corrections. The global conformal block and other contributions are shown to exponentiate in correlation functions in any dimension, corresponding to pieces of Witten diagrams at arbitrary loop order.
}

\begin{document}
\maketitle

\section{Outline}

Holographic duality \cite{hep-th/9711200,hep-th/9802109,hep-th/9802150} provides a UV complete and nonperturbative definition of a quantum theory of gravity via the boundary CFT, which offers a laboratory to shed light on long-standing questions of quantum gravity, such as the information paradox \cite{Hawking:1976ra}, and provides a rich model of emergent spacetime. While AdS/CFT has already provided much information, for example giving a model where time-evolution is manifestly unitary despite formation and evaporation of black holes, many aspects of the duality remain mysterious, and further development of the dictionary between bulk and boundary is required to realise the full potential of holographic models to shed light on quantum gravity.

A lot of progress has been made in understanding what properties of a CFT give rise to a local, semiclassical gravity dual \cite{0907.0151,1101.4163}, mostly focused on matching correlation functions to bulk perturbative effective field theory. However, the information paradox, and its sharpening by the AMP(S)S firewall argument \cite{1207.3123,1304.6483}, appear to require that bulk effective field theory should break down in some regime. An explanation of this breakdown, and of what replaces semiclassical spacetime, requires better understanding of bulk physics, and its connection to the UV description encoded in the CFT, particularly in the regime of nonlinear effective field theory.

This paper introduces a new method to organise bulk effective field theory and compute correlation functions, making use of the worldline, or `first-quantised', formalism of QFT. This is complementary to the usual Witten diagram approach, accessing slightly different physics, and having a different regime of perturbative control. Before explaining this in more detail, I highlight some aspects of the overarching motivation that are illuminated using this approach.

Best understanding of holographic duality is aided by a simple map between descriptions of bulk physics and the boundary field theory, for example decomposition of correlation functions in terms of conformal blocks, the basic kinematic ingredients of a CFT. This then allows for efficient use of field theory arguments, such as the conformal bootstrap, with a direct connection to gravitational physics. For example, 20 years after the advent of AdS/CFT, even one-loop Witten diagrams remain largely opaque, with recent progress, such as \cite{1612.03891}, based mostly on CFT crossing symmetry rather than new bulk techniques. A closely related aspect of the duality that deserves more attention is to explain the discrepancy between the natural decomposition of correlation functions from the point of view of bulk and boundary. In the bulk, a correlation function is naturally described as coming from a sum over semiclassical saddle points, with loop corrections, and depends only on the perturbative effective fields and couplings. On the other hand, in the CFT it is more natural to choose a channel, and decompose into conformal blocks in that channel, including data for all states of the theory, including those of high energy, but constrained by crossing symmetry. The approach advocated in this paper organises the bulk saddle-points in a different way from standard approaches, which gives a more direct connection between bulk and boundary than traditional Witten diagrams, and suggests new regimes in which to apply the bootstrap philosophy.

A key aspect of the bulk theory is diffeomorphism invariance, which in particular means that the bulk theory has no truly local observables. A quantum description of bulk physics therefore requires construction of appropriate diffeomorphism invariant quantities, particularly those which are sensitive to the existence of a local bulk spacetime, and which map simply to quantities in the dual field theory. Worldline-quantised particles provide natural candidates for such objects.

In the special case of three-dimensional gravity, much progress has been made by using the extended symmetries of the infinite-dimensional Virasoro algebra. There has already been much work connecting Virasoro conformal blocks to particles moving in AdS (see \cref{VirDisc} for a summary) but no systematic bulk computation of quantum corrections (though see \cref{CS} for comments on constructions in the Chern-Simons formulation \cite{1612.06385}). One could also ask how we can draw more lessons in higher dimensions from objects that seem so intimately related to the special properties of three dimensional gravity. The worldline formalism provides answers to both of these, giving a representation of Virasoro blocks with systematic corrections, which furthermore has a natural analogue in higher dimensions (though the higher-dimensional generalisation is not determined by kinematics alone).

Having made these motivating remarks, let us begin our journey into the bulk, guided by a particle leading us along its worldline.

\hrulefill

In the usual formulation of quantum field theory, the basic degrees of freedom are a set of fields $\phi(x)$, and we perform the path integral over all configurations of these fields, with some action. The worldline approach instead treats the particles as fundamental, with the configuration space being the set of particle trajectories, and the action being the length of the worldline, weighted by the particle mass. For a scalar of mass $m$ on a fixed background metric $g$, we make the following schematic replacement (in Euclidean signature):
\begin{equation}\label{2ndto1st}
	\int\displaylimits_{\substack{\text{Field}\\ \text{configurations}}} \mkern-16mu \piD{\phi}\;e^{-\frac{1}{2}\int d^{d+1}x (g^{\mu\nu}\partial_\mu\phi \partial_\nu\phi+m^2\phi^2)}\longrightarrow \int\displaylimits_{\text{Worldlines}} \mkern-12mu \piD{\gamma}\;e^{-m L[\gamma,g]}
\end{equation}
This path integral can then be cast as a one-dimensional quantum mechanics living on the worldline. As we will discuss later, once interactions are introduced, the worldline path integral reproduces Feynman rules, and thence the results of perturbative QFT.

The history of this formulation arguably goes back to Feynman \cite{Feynman:1950ir}, but interest as a practical approach began in earnest with the work of Bern and Kosower \cite{Bern:1987tw}, who computed gauge theory loop amplitudes by taking a large-tension limit of string theory. Physically, this limit causes cylinders in the worldsheet to narrow, so as to be well-approximated by particle paths, and the worldline formulation results as the QFT analogue of worldsheet string theory. In this analogy, the worldline quantum mechanics corresponds to the worldsheet CFT, and traditional QFT corresponds to string field theory. The approach used here was pioneered by Strassler \cite{hep-ph/9205205}, who used it to derive the Bern-Kosower rules directly, without recourse to string theory.

Since the worldline only gives a perturbative formulation of QFT, it may seem like a step backwards, but historically, it has proved useful for various computations beyond the applications to scattering amplitudes. Chiefly among these are applications to anomalies and index theorems, for example the early work of Alvarez-Gaum\'e and Witten computing gravitational anomalies \cite{AlvarezGaume:1983ig}. For a far more comprehensive account of the history and applications, see the review \cite{hep-th/0101036}. I will argue in this paper that the worldline gives a useful new perspective on bulk physics, from both a conceptual and technical point of view. For now, I simply outline two main points that will be central to the discussion.

Firstly, observe that the mass appears in the worldline path integral weighting the action, so $m$ plays the r\^ole of $\hbar^{-1}$. The classical limit of the worldline quantum mechanics therefore corresponds to a large mass limit, and the classical solutions are (networks of) geodesics, giving rise to the geodesic approximation to propagators and correlation functions. The loop corrections to the quantum mechanics living on the worldline then systematically correct this approximation by allowing for fluctuations of the trajectories away from geodesics, just as higher loops for worldsheet sigma-models give $\alpha'$ corrections in string theory. Conceptually, this provides a new organising principle for the bulk physics, naturally splitting contributions from different saddle-points of the worldline action, useful even outside the regime of perturbation theory. From a more pragmatic point of view, this formulation naturally lends itself to systematic calculations in the perturbative regime of large mass (in AdS units). We will find that this is complementary to the usual bulk perturbation theory, and accesses different physics, for example allowing for simple computations of contributions that would be at arbitrarily high loop order in the usual Witten diagram expansion.

Secondly, it is simple to couple the worldline theory to external fields such as gravity, as in \cref{2ndto1st} by including the metric dependence on the worldline length. We can then quantise the external field by the standard approach, integrating over configurations, and in the theory of the external field, the worldline path integral appears as a nonlocal operator:
\begin{equation}
	\int\piD{g}\;e^{-\frac{1}{G_N} S_{EH}[g]}\int\piD{\gamma} \;e^{-m L[x,g]}
\end{equation}
A standard example of this is the integral over closed worldlines of a charged particle coupled to an electromagnetic field, which, from the point of view of the Maxwell field, generates a series of interaction vertices. These are interpreted as the irrelevant operators generated by integrating out the charged particle at one (or more) loop, deforming QED to include photon vertices, for example allowing scattering of light by light via an electron in a loop. For calculations perturbative in the strength of the external field or its coupling, the interaction with the worldline can then be put in terms of vertex operators, whose correlation functions we can calculate in the worldline quantum mechanics.\footnote{In string theory, the vertex operators have a dual r\^ole to play, since external fields which we can couple to map to the states of the string itself. This needs conformal invariance on the worldsheet and the state-operator correspondence, so there is no analogue here.} Applied to AdS/CFT, this will give us a second perturbation expansion (on top of the loop expansion in $m^{-1}$) in the strength of the external field, parametrised by $G_N$.
 
Keeping these two conceptual points in mind, I now outline the main context in which we will apply them. Consider, for definiteness, a CFT in $d$ dimensions with a semiclassical, local AdS$_{d+1}$ dual (with curvature length scale $\ell_{AdS}$), described at low energies by Einstein gravity coupled to matter, with Planck scale parametrically separated from the AdS scale ($G_N\ll \ell_{AdS}^{d-1}$). Particularly if the matter is heavy, which here means $m\ell_{AdS}$ is large (but $m^{d-1} G_N\ll 1$ to keep control of gravitational effective field theory), it is most natural to quantise it in the worldline formalism. A physical way of saying this is that the Compton wavelength of the particle is intermediate between the Planck and AdS scales, which means that for AdS scale physics the appropriate classical limit is described in terms of particles and not fields. The quantum mechanics on the worldline is then approximately in a classical limit. While it is simplest to keep this regime in mind, I emphasise here that many of the lessons learnt will also apply at finite $m\ell_{AdS}$ (or alternatively at Planckian masses, particularly for $d=2$).

We therefore quantise the `heavy' matter using the worldline formalism, while using the standard QFT approach for the `light' fields, in particular gravity (though more generally we could include gauge fields or anything else in this category). A heavy scalar particle is dual to a single-trace scalar primary operator $\phi(x)$ in the CFT, with dimension $\Delta\gg 1$ in the perturbative regime of the worldline quantum mechanics. The specific quantity that we will be most interested in is the four-point function $\langle\phi(x_1)\phi(x_2)\phi(x_3)\phi(x_4)\rangle$ of this heavy operator, which is computed by the path integral with boundary conditions that there are worldlines ending on the boundary at the insertion points $x_i$. The most important classical solutions with these boundary conditions involve a pair of disjoint worldlines, on geodesics in pure AdS, joining the points $x_1,x_2$ and $x_3,x_4$, for example. This is not in fact a full classical solution, since the worldlines should backreact on the geometry, but this backreaction is parametrically small so we can nonetheless use it as a starting point for perturbation theory. This perturbation theory is in terms of two small parameters, $m^{-1}$ and $mG_N$, taking into account the fluctuations of the worldline away from the geodesic classical solution, and the exchange and bulk interaction of gravitons respectively.

The main work of the paper is to set up this expansion in detail, resulting in a systematic perturbation theory. The focus is primarily on the conceptual, and I will compute only some of the simplest contributions to this expansion, leaving more detailed technical work for future development. Despite this, even the simple diagrams computed here will have interesting and novel consequences, exemplifying the utility of the approach. I now give a brief summary of the main results in this direction. I state these in the language of a heavy scalar coupled to gravity, but in almost all cases, analogous results apply with more general field content.

\begin{itemize}
	\item To first nontrivial order in $G_N$, the worldline formalism gives the global conformal block corresponding to the exchange of the stress tensor and descendants.
	\item In doing so, the `geodesic Witten diagram' prescription \cite{1508.00501} emerges naturally; this could be regarded as an explanation of why that recipe works.
	\item The stress tensor block, and all higher contributions, exponentiate in the correlation function. This is trivial to see from the worldline perspective, but unexpected, with far-reaching consequences, from standard CFT or Witten diagram points of view, in particular encoding contributions to Witten diagrams at arbitrary loop order, and to the OPEs of high multi-trace operators.
	\item As is evident from the conformal block result, the worldline naturally and cleanly splits up contributions corresponding to different particle number. In particular, the single- and double-trace exchanges in a tree level exchange Witten diagram come from different saddle points in the worldline path integral.
	\item In AdS$_3$, the result to all perturbative orders is the Virasoro conformal block. In an appropriate classical limit, this is a `semiclassical block', with the worldline path integral giving a systematic bulk method to compute and interpret quantum corrections to any desired order.
	\item Combining the two results above, it becomes clear that the Virasoro conformal block does not include the full contribution to a correlation function from gravitational interactions. Gravitational dynamics in AdS$_3$ is not fixed completely by kinematics, when coupled to matter.
	\item Again special to the theory in AdS$_3$, the quantum mechanics on the worldline is one-loop exact, computable by fermionic localisation (though I do not exploit or develop this observation further here).
\end{itemize}

From the point of view of understanding worldline QFT, some of the ideas in section 3 are, as far as I am aware, new. In particular, the boundary conditions with the worldlines going to infinity, as required to compute their correlation functions, is different from most of the worldline QFT literature, which deals with loops of the first-quantised particle. Consideration of this leads to new results that are applicable more generally. Working in an AdS background specifically is also new, though is largely an application of the standard formalism.

\hrulefill

The paper is structured as follows. In \cref{diagrammatica}, I outline the main idea, sketching the diagrammatic organisation of the two-parameter perturbation theory resulting from coupling gravity to scalars quantised in the worldline formalism. In \cref{worldlineSec}, I begin to set up the details, reviewing the worldline analogue of the Polyakov path integral, describing the quantum mechanics that lives on the worldline, its perturbative expansion, and the coupling to gravity. In \cref{4pt} I then apply this to the main example of the paper, the four-point correlation function of scalar operators, in which most of the results previewed above will be explained. \Cref{disc} expands on the discussion in various directions, including sketching how the ideas apply to different backgrounds, focussing on thermal AdS, and a more detailed discussion of the connection to Virasoro conformal blocks and the existing literature on that subject. I also briefly discuss various generalisations and future directions of study, including Lorentzian and other special kinematics, and the connection to constructions of diffeormorphism invariant bulk observables.

\section{Perturbation theory for worldlines coupled to gravity}\label{diagrammatica}
% #ArsDiagrammatica	#Diomedes #Latin #ClassicalEducation #Puns

In this section, I will sketch the general idea, outlining the structure of perturbation theory in the simplest context of Einstein gravity coupled to a worldline-quantised scalar particle, focussing for definiteness on the four-point function of a heavy operator $\phi$ as described in the introduction. This section will not use any details particular to AdS backgrounds, and is straightforwardly generalised to other field content and couplings, I simply choose one example to streamline the exposition. The technical details will be developed in subsequent sections.

The worldline formulation treats the particle dual to $\phi$ as fundamental rather than the field, quantising the classical theory of a relativistic particle by integrating over all trajectories through spacetime. This requires choosing a parametrisation of this configuration space, gauge-fixing, finding the correct measure for the path integral and so forth, discussed in the next section, but here we focus simply on the structure of the perturbation theory. For this purpose, it will be enough to write down the schematic form of the path integral over worldlines with some action $S_{WL}$, depending on the particle path and the metric, as well as over metrics to quantise gravity in the usual way, with action $S_{EH}$:
\begin{equation}
	\int\piD{g}\piD{\gamma}\;e^{-\frac{1}{G_N} S_{EH}[g]-m S_{WL}[\gamma,g]}
\end{equation}
It is simplest to consider the parametric regime $1\ll m \ll G_N^{-1}$ (here and henceforth we set $\ell_{AdS}=1$), in which the natural way to compute the path integral, with appropriate boundary conditions for the metric and worldlines, is as follows. First, find the saddle-points $g=g_0$ of the dominant Einstein-Hilbert piece of the action $S_{EH}[g]$, that is solving the vacuum Einstein equations with the given boundary conditions. Then, find saddle points $\gamma=\gamma_0$ of the worldline action on this fixed background $S_{WL}[\gamma,g_0]$, which are geodesics in $g_0$. The result is not a true stationary point of the full action, since we have not accounted for the backreaction of the particle, but an approximate one if $m G_N\ll 1$ (a `probe limit'). Finally, compute perturbative corrections in powers of $G_N$ and $m^{-1}$ from the fluctuations of metric and particle path about the saddle point ($g=g_0+h$ and $\gamma=\gamma_0+q$ with $h$ and $q$ small), and sum over contributing saddles. There may be several separate worldlines included in the classical solution, perhaps joined at vertices, included to account for interactions of heavy particles. For the four-point function $\langle\phi\phi\phi\phi\rangle$, the most important saddle-point takes $g_0$ to be pure AdS, with two separate geodesics, each joining a pair of the operator insertion points on the boundary.

The natural way to describe the particle path integral is in terms of fluctuations of the particle trajectory away from the geodesic. These are described by fields, collectively denoted $q$, that live on the worldline, depending only on a single `time' coordinate, such as an affine parameter $s$ for the geodesic. The worldline is then parametrised as $x^a(s)=x_0^a(s)+q^a(s)$, where $x^0$ is the geodesic path, and we integrate over the $q$ fluctuations. Their contribution can be therefore described by a quantum mechanics living on each geodesic segment, which we will explore in more detail in the next section. This quantum mechanics may be exactly solvable, but in most cases it will be more convenient to expand order by order in the fluctuations $q$ and compute perturbatively. The coupling constant in this expansion is roughly the inverse mass of the particle in AdS units $m^{-1}$, with fluctuations exploring AdS away from the geodesic suppressed by the Compton wavelength of the particle in units of the curvature scale.

We also expand in metric fluctuations $h$. From the point of view of the worldline, $h$ is regarded as an external field, and expanding $e^{-S_{WL}}$ in powers of $h$ will result in factors of integrals of local operators of the quantum mechanics, the vertex operators $\mathcal{V}$.
The path integral over worldlines with the external field $h$ turned on gives, at each order in metric perturbations, the correlation functions of products of vertex operators in the quantum mechanics. Finally, for the bulk theory itself we have the usual perturbative expansion, with graviton propagators of order $G_N$ and vertices of order $G_N^{-1}$, with the only novelty that the coupling to the worldline gives sources in the bulk.

We can cast this in diagrammatic language, with the quantum mechanics giving propagators and interactions corresponding to fluctuations $q$, living on a particular segment of worldline, gravity having its usual graviton propagators and vertices in the bulk, and the vertex operators providing couplings between $q$ and $h$, of order $m$. Note that because we have not solved the Einstein equations exactly, but only in the probe limit, these include couplings of the graviton directly to the worldline (a classical source for $h$), $\mathcal{V}$ including a piece independent of $q$. This means that there will be vertices with gravitons coupling to any number of $q$ propagators, including zero and one. We sketch the Feynman rules, including only their scaling with the parameters, for the worldline theory, the bulk theory, and their couplings, in \cref{FeynmanRules}.
\begin{table}
	\centering
	\begin{subtable}{.3\textwidth}
	\centering
	\begin{tabular}{Vl}
		\begin{tikzpicture}
		\path (0,.9)--(0,-.7);
		\draw  (0,0) -- (1.3,0);
	\end{tikzpicture} & $\sim m^{-1}$ \\
		\begin{tikzpicture}
		\path  (0,.9)--(0,0);
		\draw  (0,0) -- (-1,0);
		\draw  (0,0) -- (.5, {sqrt(3)/2});
		\draw  (0,0) -- (.5, {-sqrt(3)/2});
	\end{tikzpicture} & $\sim m$ \\
	\begin{tikzpicture}
		\path (0,1)--(0,0);
		\draw (.7,.7) -- (0,0);
		\draw (.7,-.7) -- (0,0);
		\draw (-.7,.7) -- (0,0);
		\draw (-.7,-.7) -- (0,0);
	\end{tikzpicture} & $\sim m$
	\end{tabular}
	\end{subtable}
	\begin{subtable}{.3\textwidth}
	\centering
	\begin{tabular}{Vl}
		\begin{tikzpicture}
		\path (0,.9)--(0,-.7);
		\draw [grav] (0,0) -- (1.3,0);
	\end{tikzpicture} & $\sim G_N$ \\
		\begin{tikzpicture}
		\path  (0,.9)--(0,0);
		\draw [grav]  (0,0) -- (-1,0);
		\draw [grav] (0,0) -- (.5, {sqrt(3)/2});
		\draw [grav] (0,0) -- (.5, {-sqrt(3)/2});
	\end{tikzpicture} & $\sim G_N^{-1}$ \\
	\begin{tikzpicture}
		\path (0,1)--(0,0);
		\draw [grav]  (.7,.7) -- (0,0);
		\draw [grav] (.7,-.7) -- (0,0);
		\draw [grav] (-.7,.7) -- (0,0);
		\draw [grav] (-.7,-.7) -- (0,0);
	\end{tikzpicture} & $\sim G_N^{-1}$ \\
	\end{tabular}
	\end{subtable}
	\begin{subtable}{.3\textwidth}
	\centering
	\begin{tabular}{Vl}
		\begin{tikzpicture}
			\path (0,.8)--(0,-.6);
			\node [draw,thick,circle,cross,inner sep=0, minimum size=7](A) at (0,0){};
			\draw [grav] (A) -- (1.1,0);
		\end{tikzpicture} & $\sim m$ \\
		\begin{tikzpicture}
			\path (0,.8)--(0,-.6);
			\node [draw,circle,fill=black,inner sep=0pt,minimum size=3pt] at (0,0){};
			\draw (0,0) -- (-.5,0);
			\draw [grav] (0,0) -- (.6,0);
		\end{tikzpicture} & $\sim m$ \\
		\begin{tikzpicture}
			\draw [grav]  (0,0) -- (-1,0);
			\draw (0,0) -- (.5, {sqrt(3)/2});
			\draw (0,0) -- (.5, {-sqrt(3)/2});
		\end{tikzpicture} & $\sim m$
	\end{tabular}
	\end{subtable}
	\caption{\label{FeynmanRules}Parametric scaling of vertices and propagators in Feynman rules. The worldline theory is in the first column, with solid lines corresponding to fluctuations $q$ of the worldline. The second column shows the rules for the usual gravitational theory in the bulk. The third column gives the couplings between the two. Higher-point vertices scale in the same way. The vertex at the top right, indicating a graviton coupling directly to the worldline without any fluctuation, exists because we are perturbing around a probe limit, in which we have not solved the full equations of motion including backreaction of the particle.}
\end{table}

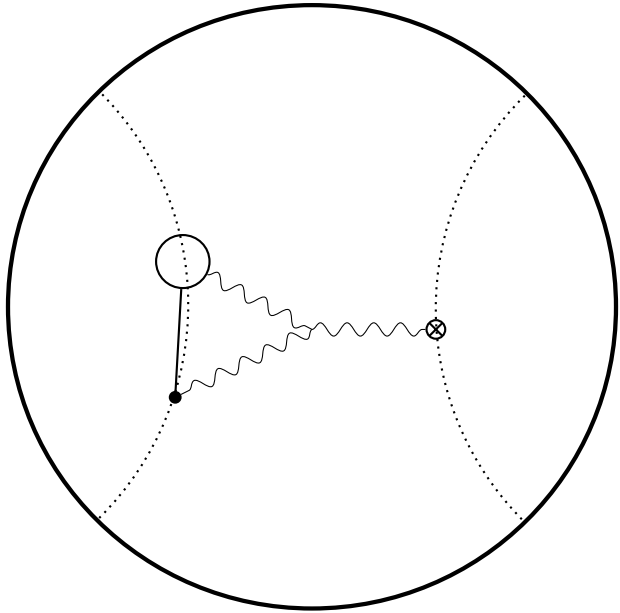
\begin{figure}
	\centering
	\begin{tikzpicture}
%		\draw [help lines] (0,0) grid (8,8);
		\draw [ultra thick] (4,4) circle(4);
		\draw [thick, dotted] (1.2,1.2) arc (-45:45:4);
		\draw [thick, dotted] (6.8,6.8) arc (135:225:4);
%		\draw [thick] (2.3,4.4) circle(.3);
%		\draw [thick] (5.75,3.1) circle(.3);
%		\draw [grav] (2.57,4.3) -- (5.45,3.2);
		\node [draw,thick,circle,cross,inner sep=0, minimum size=7](A) at (5.63,3.7){};
		\node [draw,thick,circle,fill=black,inner sep=0pt,minimum size=4](B) at (2.2,2.8){};
		\node [draw,thick,circle,minimum size=20](C) at (2.3,4.6){};
		\coordinate (int) at (4,3.7);
		\draw [grav] (C) -- (int);
		\draw [grav] (int) -- (B);
		\draw [grav] (int) -- (A);
		\draw [thick] (B) --(C);
	\end{tikzpicture}
	\caption{A two-loop, two graviton diagram, of order $m G_N^2$. The dotted lines indicate the locations of the unperturbed geodesics for the two particles, with an independent quantum mechanics living on each of these. This diagram contributes to $\langle \langle\mathcal{V}_h\mathcal{V}_h\rangle_{WL1}\langle\mathcal{V}_h\rangle_{WL2}\rangle_\text{gravity}$, with a one-loop contribution to the two-point function of the graviton vertex operator on the first worldline, the tree-level one-point function of the vertex operator on the second, and a gravity three-point function of $h$ to tie them all together.\label{exampleDiagram}}
\end{figure}

With these rules in hand, what should we compute? The perturbative corrections to the background (that is, the metric along with the geodesics) are encoded as the sum of vacuum diagrams with these rules, with no external sources. In the usual way, this is the exponential of the sum of connected vacuum diagrams, a simple fact that will have nontrivial consequences when applied later. The connected diagrams have a two-parameter expansion, in $m^{-1}$ and $G_N$, with the classical action of the geodesics contributing at order $m$, and the diagrams providing successively smaller corrections to that, with finitely many diagrams at any given order. It is convenient to organise this expansion in powers of $m^{-1}$ and $m G_N$, which respectively count the number of loops $L$ in the diagram, and the number of gravitons $N_g$ exchanged (roughly speaking), to give an overall contribution proportional to $m^{1+N_g-L}G_N^{N_g}$. As an example, a two-loop, two-graviton diagram is shown in \cref{exampleDiagram}. The loops can be thought of as corrections to the geodesic approximation, and the gravitons correct the probe approximation. The main result of the paper is to set up a systematic expansion of these corrections.

If we wish to include interactions of the first-quantised particles by starting with a network with geodesics meeting at vertices, we should additionally allow the fluctuations $q$ living on different geodesic segments to interact at the vertices, coupling the boundary conditions by imposing that these interactions conserve bulk momentum. I will not develop this in any detail here, but it does not add any new conceptual difficulties.

I conclude the section by comparing with the usual Witten diagram expansion. Since the number of gravitons $N_g$ counts the power of $G_N$ in a diagram, it may appear that $N_g-1$ is equivalent to the loop order of a corresponding Witten diagram to which it contributes, but this is not quite correct, because the connected diagrams exponentiate. Any given connected worldline diagram encodes some piece of infinitely many Witten diagrams, to arbitrarily high loop order. Obversely%	This is an excellent word. You should use it, dear diligent source-reader. No, I don't mean conversely.
, any single Witten diagram requires an all-loop (and, as we will see later, even nonperturbative) calculation on the worldline to reconstruct it. The standard approach and that advocated in this paper are therefore complementary, each being more useful and natural in its own appropriate regime.

\section{The worldline quantum mechanics \label{worldlineSec}}

The previous section outlined the general structure of the perturbation theory of a first-quantised scalar particle coupled to gravity. In this section, we will begin to attack the details, by learning more about the quantisation of the worldline theory. Some of this will be a lightning review of worldline quantum field theory, which is supplemented by \cref{WLreview}, though for more detail and references, see \cite{hep-th/0101036,Bastianelli:2006rx}. The main conceptually new material is in \cref{BCinf}, describing how boundary conditions at infinity simplify the worldline path integral. The section develops all the details of the worldline theory in AdS, culminating in the Feynman rules for the perturbative computation of the vertex operator correlation functions, carried out in \cref{VOcorrelators}.

The details of the worldline theory are not necessary to understand the main points of the following section, so a reader impatient to apply the formalism to AdS correlation functions could skip to the next section.

\subsection{Gauge-fixed worldline path integral}

The first thing to do to make the worldline formulation more concrete is to define what is meant by the path-integral
\begin{equation}
	\int \piD{\gamma}\; e^{-m L[\gamma,g]}
\end{equation}
where the integral is over all paths, with some boundary conditions (either fixed endpoints, or periodic). A path is described by a map $x^a(s)$ from an interval (or a circle for periodic boundary conditions) parametrised by $s$, to spacetime, with coordinates $x^a$, so we can integrate over all such functions, with the length $L=\int ds \sqrt{g_{ab}(x)\dot{x}^a\dot{x}^b}$. But this overcounts, because the same path can be parametrised in many ways: there is a gauge symmetry of diffeomorphisms on the worldline that we must fix.

The standard way to proceed is to introduce an auxiliary metric on the worldline, described by an `einbein' $e(s)=\sqrt{g_{ss}(s)}$ such that $e(s)ds$ defines the length element, with the action
\begin{equation}
	\quad S_P[x,e,g]=\frac{m}{2}\int ds\, e(s) \left(e(s)^{-2}g_{ab}(x(s))\dot{x}^a(s)\dot{x}^b(s) + 1 \right),
\end{equation}
in which we can replace $e$ by the solution to its equation of motion to show equivalence to the usual length functional. Now for any given $e$, we can choose $s$ to be the arc-length along the curve with respect to this auxiliary metric, or said another way, there is always a diffeomorphism that sets $e(s)=1$. This condition defines the commonly-used gauge in worldline QFT, which I will call \emph{Polyakov gauge}, after its introduction in Polyakov's book \cite{Polyakov:1987ez}, and its close analogue to the Polyakov path integral for the string. In that context, we introduce an auxiliary worldsheet metric, use Weyl gauge transformations to fix that metric to be flat (locally), and then diffeomorphisms to choose standard coordinates.

Just as for the string, there is global gauge-invariant data on the worldline that we can't alter by diffeomorphisms, leaving a residual integral over `moduli'. For a particle with boundary conditions of fixed initial and final endpoints $x_0,x_1$, the parameter is defined on an interval, and the modulus is just the length $T$ of the worldline as measured in the auxiliary metric, $T=\int e(s) ds$. After fixing the starting point to be at $s=0$, the action integral is between $s=0,T$, and integrating over $T$ finally gives the gauge-fixed path integral:
\begin{equation}\label{WLPI}
	\int \piD{\gamma}\; e^{-m L[\gamma,g]}
\longrightarrow \frac{1}{2m}\int_0^\infty dT e^{-\frac{m T}{2}} \int\limits_{\substack{x(0)=x_0 \\x(T)=x_1}} \mkern-16mu \piD{x} \;e^{-\int_0^T ds \frac{m}{2} g_{ab}\dot{x}^a\dot{x}^b}
\end{equation}
In \cref{polyakov} I review the more careful derivation of this result by the Fadeev-Popov procedure. The measure from gauge-fixing does not introduce any subtleties, with the ghosts providing only a trivial theory which decouples. The prefactor is a conventional normalisation, chosen for later convenience.

We will also on occasion be interested in the case when the worldline forms a closed loop, the more commonly seen case in the literature. Now again we cannot fix the proper time around the loop by gauge transformations, so must integrate over it, but there is now an ambiguity over where to put $s=0$ on the circle (or alternatively, constant shift diffeomorphisms are left unfixed). To compensate for the resulting overcounting, we can divide by $T$ in the measure for the modulus:
\begin{equation}\label{worldlineZ}
	\frac{1}{2m}\int_0^\infty \frac{d T}{T} e^{-\frac{m T}{2}} \mkern-16mu\int\displaylimits_{x(0)=x(T)}\mkern-16mu \piD{x} \;e^{-\int_0^T ds \frac{m}{2} g_{ab}\dot{x}^a\dot{x}^b}
\end{equation}

The worldline path integral over $x$ we are left with is something very familiar, defining a quantum-mechanical sigma-model, with spacetime as the target space. In particular, we can define the path integral to reproduce the results obtained from a Hamiltonian proportional to the Laplacian on the target space. The path integral with fixed boundary conditions on the endpoints is then given by the heat kernel (the heat equation being the Euclidean Schr\"odinger equation):
\begin{equation}
	\int\limits_{\substack{x(0)=x_0 \\x(T)=x_1}} \mkern-16mu \piD{x} \;e^{-\int_0^T ds \frac{m}{2} g_{ab}\dot{x}^a\dot{x}^b} = K(x_0,x_1;T),\quad \frac{\partial K}{\partial T} = \frac{1}{2m}\nabla_g^2 K,\quad K(x_0,x_1;0)=\delta(x_0,x_1)
\end{equation}
For the particle on an interval, the integral over $T$ gives the Schwinger or heat kernel representation of the Klein-Gordon propagator (formally, writing $K$ as a matrix element of the operator $\exp\left(\frac{T}{2m}\nabla^2\right)$, the integral gives $\frac{1}{m^2-\nabla^2}$). Gluing worldline path integrals together at vertices thence recovers the standard position-space Feynman rules (or Witten diagrams in AdS). Similarly, the integral over closed loops is interpreted as the heat kernel regularisation (after cutting off the modulus integral at small  $T$) of the one-loop partition function of a free massive scalar.

\subsection{Perturbation theory on the worldline\label{WLpert}}

We now wish to apply this to the calculations described in \cref{diagrammatica}, computing correlation functions of vertex operators associated to gravitons or other external fields, in perturbation theory on the worldline. 

In the classical limit, the sigma-model is governed by a stationary point of $\int g_{ab}\dot{x}^a\dot{x}^b$, which is an affinely parameterised geodesic. The normalisation of the parameter is fixed by the boundary conditions to be $\frac{T}{L}$ times the arc length along the geodesic, where $L$ is the total proper length from $x_0$ to $x_1$. It is convenient to reparameterise by absorbing this factor of $\frac{T}{L}$ into the parameter $s$, so that the action appearing in the exponential is now given by
\begin{equation}\label{action}
	S[x] = \frac{1}{2\lambda} \int_0^L ds\; g_{ab}\dot{x}^a\dot{x}^b, \quad \text{with `coupling'}\quad \lambda=\frac{T}{mL}.
\end{equation}
It is now clear that the amplitude does not depend separately on $T$ and $m$, but only through $\lambda$. The theory is weakly coupled, and hence admits a perturbative expansion, when $\lambda$ is small, as measured in units of the typical curvature of the target manifold.

From this, the integral over the modulus $T$ is an integral over the coupling of the sigma model, which might seem like a problem if we are able to compute only at small $\lambda$. In fact, at large mass, we do not need the whole integral, since there is a saddle-point in $T$ centred at weakly coupled values. In the classical limit, the amplitude is dominated by the on-shell action $\frac{L}{2\lambda}=\frac{mL^2}{2T}$, suppressed at small $T$, while the factor $e^{-mT/2}$ in \cref{WLPI} is suppressed at large $T$, and the combination of these terms results in an integral over moduli dominated by the saddle point where $\frac{mT}{2}+\frac{mL^2}{2T}$ is minimised at $T=L$.
Including loop corrections to the path integral, we can simply expand in the saddle-point approximation around this point. This approximation is valid if $mL\gg 1$, and as long as the loop corrections are small at coupling $\lambda\approx m^{-1}$, which requires $m$ to be large in units of the curvature of the target space.

To illustrate this explicitly, consider the main example of the paper, with the target space AdS$_{d+1}$. A convenient set of coordinates are those of static global AdS, usually written in terms of $t,r$ and angles on $S^{d-1}$, but it is useful here to exchange the polar coordinates $r$ and the angles in favour of `Cartesian' coordinates $q^i$, $i=1,2,\ldots,d$:
\begin{equation}
	ds^2 = (1+q^2)dt^2 +dq^2 -\frac{(q\cdot dq)^2}{1+q^2}.
\end{equation}
Here, the indices of $q$ are contracted with $\delta_{ij}$ so, for example, $q^2=\sum_i q^i q^i$, and $q\cdot dq$ denotes $\sum_i q^i dq^i$. For any given geodesic, we can always choose these coordinates such that it lies along the origin $q=0$, and the isometries that map the geodesic to itself are the manifest $\RR\times SO(d)$ subgroup of the full $SO(d+1,1)$ isometry group.

Now to describe the AdS propagator at proper distance $L$, we perform the worldline path integral with endpoints located at $t=0,q=0$ and $t=L,q=0$. Writing $t(s)=s+u(s)$ so that $u(s)$ and $q^i(s)$ are fluctuations away from the geodesic, with boundary conditions $u(0)=q(0)=u(L)=q(L)=0$, the action is
\begin{equation}
	S[x] = \frac{L}{2\lambda} + \frac{1}{2\lambda}\int_0^L ds \left(\dot{u}^2+\dot{q}^2+q^2 +2\dot{u}q^2+\dot{u}^2 q^2 - \frac{(q \cdot\dot{q})^2}{1+q^2} \right).
\end{equation}
The quadratic piece of the action is given by a massless field $u$, which represents unphysical fluctuations of the parametrisation, along with $d$ simple harmonic oscillators $q^i$ of frequency $\ell_{AdS}^{-1}$, the physical transverse oscillations of the particle. The classical limit gives rise to the theory of a nonrelativistic particle moving near the centre of AdS, where the important effect is the quadratic gravitational potential, leading to the theory of $d$ harmonic oscillators. The remaining terms give couplings between the fields, and at this point it is straightforward to write Feynman rules and start computing order by order in a loop expansion in $\lambda$, in particular expanding the final term to give couplings between $q$s of all even valencies, $4,6,8,\dots$. This gives a diagrammatic way to compute the short-time asymptotic expansion of the heat kernel.

The Feynman rules for these boundary conditions, with finite $L$, are written in \cref{HKQM}, where I provide details of the two-loop calculation of the partition function (the simpler rules for $L\to\infty$ will appear in the next section), with the following result:
\begin{align}\label{twoLoopWL}
	W_L(\lambda)&=-\log\int \piD{x} \;e^{-S[x]}\\
	 &= S_\text{on-shell} -\left[ \begin{tikzpicture}[baseline={([yshift=-.5ex]current bounding box.center)}]
			\node[draw,circle,minimum size=15] at (0,0){};
		\end{tikzpicture}
		+ \begin{tikzpicture}[baseline={([yshift=-.5ex]current bounding box.center)}]
			\node[draw,dashed,circle,minimum size=15] at (0,0){};
		\end{tikzpicture}
		+ \begin{tikzpicture}[baseline={([yshift=-.5ex]current bounding box.center)}]
			\node[draw,circle,minimum size=15](A) at (0,0){};
			\node[draw,circle,minimum size=15](B) at (1,0){};
			\draw[densely dashed] (A) --(B);
		\end{tikzpicture}
		+ \begin{tikzpicture}[baseline={([yshift=-.5ex]current bounding box.center)}]
			\draw (0,0) circle(.3);
			\draw[densely dashed] (-.3,0) -- (.3,0);
		\end{tikzpicture}
		+ \begin{tikzpicture}[baseline={([yshift=-.5ex]current bounding box.center)}]
			\draw (0,0) to[out=45,in=90, looseness=1] (.8,0) to [out=-90,in=-45, looseness=1.5] (0,0);
			\draw (0,0) to[out=135,in=90, looseness=1] (-.8,0) to [out=-90,in=-135, looseness=1.5] (0,0);
		\end{tikzpicture}
		+ \begin{tikzpicture}[baseline={([yshift=-.5ex]current bounding box.center)}]
			\draw[densely dashed] (0,0) to[out=45,in=90, looseness=1] (.8,0) to [out=-90,in=-45, looseness=1.5] (0,0);
			\draw (0,0) to[out=135,in=90, looseness=1] (-.8,0) to [out=-90,in=-135, looseness=1.5] (0,0);
		\end{tikzpicture}
		 +\cdots\right]
		 \nonumber\\
	 &= \frac{L}{2\lambda}+\frac{d}{2}\log\left(2\pi\lambda\sinh L\right)+\frac{1}{2}\log\left(2\pi\lambda L\right)+\left[\frac{d(d-2)}{8}\left(\frac{1}{L}-\frac{1}{\tanh L}\right)+\frac{d^2}{8}L\right]\lambda + O(\lambda^2)\nonumber
\end{align}
The one-loop contributions to $\Gamma$ are simply the logarithms of the quantum mechanics propagators of a free particle, for $u$, and $d$ harmonic oscillators of unit frequency and mass $\lambda^{-1}$, for $q^i$. The result reproduces the correct short-time asymptotic expansion of the heat kernel in hyperbolic space (for which, see \cref{largedim}).

With this in hand, we simply need to perform the Laplace transform in \cref{WLPI} to find the propagator, with $\lambda=\frac{T}{mL}$:
\begin{align*}
	\frac{1}{2m}\int_0^\infty & dT e^{-\frac{mT}{2}-\frac{mL^2}{2T}}\left(\frac{m}{2\pi T}\right)^\frac{d+1}{2} \left(\frac{L}{\sinh{L}}\right)^\frac{d}{2}\left[1-W_L^{(2)}\frac{T}{mL}+O(m^{-2})\right] \\
	= \frac{e^{-m L}}{2m}\left(\frac{m}{2\pi \sinh L}\right)^{d/2}&\int  \frac{dx}{\sqrt{2\pi}}e^{-\frac{x^2}{2}+\frac{x^3}{2\sqrt{mL}}-\frac{x^4}{2mL}+\cdots}\left(1+\frac{x}{\sqrt{mL}}\right)^{-\frac{d+1}{2}}\left(1-W_L^{(2)} \frac{1+\frac{x}{\sqrt{mL}}}{m}+\cdots \right)\\
	&= \frac{e^{-m L}}{2m}\left(\frac{m}{2\pi \sinh L}\right)^{d/2}\left(1+\frac{d(d-2)}{8m\tanh L}-\frac{d^2 L}{8m}+\cdots \right)
\end{align*}
Here in the second line I have written $T=L+\sqrt{\frac{L}{m}}x$, before expanding the integrand in powers of $m$ and doing the $\int_{-\infty}^\infty dx\; e^{-\frac{x^2}{2}}x^n$ integrals to get the result in the final line ($W_L^{(2)}$ denotes the two-loop contribution to $W_L(\lambda)$, the expression in the square brackets in the final line of \cref{twoLoopWL}). This matches a large-mass expansion of the AdS propagator, computed directly in \cref{largedim}.

Notice that the validity of the saddle-point approximation made here requires that the width of the saddle-point is small, demanding $mL\gg 1$. I have introduced this in the context of fixed $L$ and large $m$, but it seems from this that it may be applicable in the alternative limit of fixed $m$ and large $L$. This is nearly true, with the caveat that the loop expansion may not be valid, the terms in $W_L(\lambda)$ growing linearly in $L$ in particular being important. But this merely shifts the position of the saddle-point, and the integral over $T$ still concentrates on a single value of the modulus. I will develop this in more detail in the next subsection.

Before turning to this large $L$ limit, I should first highlight some subtleties with the perturbation theory of the sigma-model path integral that have been swept under the rug thus far.
Since we are describing a one-dimensional quantum mechanics, furthermore in which the Hamiltonian has such a simple and explicit description as the Laplacian, one might na\"ively think that the path integral and perturbation theory would be straightforward to define and work with, but in fact one must be careful.

Firstly, the measure integrating over paths $x(s)$ must be chosen to preserve bulk diffeomorphism invariance, in particular respecting the proper volume form. Schematically writing the path integral as a discretised product over successive points, the measure should look like
\begin{equation}
	\piD{x} \approx \prod_s dx^0(s) dx^1(s)\cdots dx^d(s) \sqrt{\det g(x(s))}\;,
\end{equation}
with field-dependence introduced through the determinant of the metric. Without this insertion, the path integral is not even finite, despite the theory being one-dimensional. For the AdS path integral in the coordinates used here, we in fact avoid the subtlety, since $\det g$ is unity, but once we perturb the metric by turning on gravity, this will become important. The determinant can be accounted for by introducing additional ultralocal ghost fields \cite{hep-th/9112035,hep-th/0101036,Bastianelli:2006rx}, though since in this paper I will only describe fluctuations around AdS, I will be able to avoid this, including the correct measure in an alternative way when required.

Even having done this, applying the Feynman rules results in ambiguous integrals of products of distributions, and regularising the path integral in different ways gives different answers. This is well-discussed and explored in \cite{hep-th/9504097}, to which I refer the interested reader; I here briefly summarise the upshot. Ambiguities in operator ordering of the Hamiltonian, and discretisation of the path integral, break manifest covariance, and the correct result reproducing the Hamiltonian proportional to $\nabla^2$ requires the addition of a non-covariant quantum correction to the action \cref{action}, $S_q[x]=-\frac{\lambda}{8} \int_0^L ds (R+g^{ab}\Gamma^c_{\phantom{a}ad}\Gamma^d_{\phantom{a}bc})$.\footnote{This counterterm is regularisation dependent, the one given relevant for the `time-slicing' regulator \cite{hep-th/9504097}.} With the extra quantum terms included, and a careful treatment of the path integral, the ambiguous products of distributions can be given definite values, which correctly reproduce the heat kernel. Again, for pure AdS in the coordinates used here, the correction term is relatively unimportant, since $\Gamma\Gamma$ vanishes, and $R$ is just a constant, $R=-d(d+1)$. However, doing calculations of graviton vertex operators at sufficiently high order requires both terms to be accounted for. In fact we have already used the $R$ correction term in the two-loop calculation \cref{twoLoopWL} above, to get the correct coefficient of $L\lambda$, as explained in more detail in \cref{HKQM}.

Finally, I comment on the special case of AdS$_3$, for which the reader may notice that the two-loop correction in \cref{twoLoopWL} vanishes (apart from a shift in the ground state energy). In fact, this happens for a good reason, that the worldline sigma-model is one-loop exact, from the Duistermaat-Heckman theorem \cite{duistermaat1982variation}. For the positively-curved cousin $S^3$ of AdS$_3$, this follows immediately from the fact that the target space is the $SU(2)$ group manifold \cite{Picken:1988ev}, though the same conclusion holds for hyperbolic 3-space (Euclidean AdS$_3$). From a physics perspective, this can be shown from fermionic localisation techniques (despite the model being purely bosonic), reviewed in \cite{hep-th/9608068} (see also \cite{1703.04612} for a nice review, and recent application of the same techniques to the Schwarzian theory). It would be interesting to develop the details of this, and leverage it for computation of vertex operator correlation functions, but I leave this exploration to future work.

\subsection{Boundary conditions at infinity \label{BCinf}}

The discussion above describes the worldline path integral with fixed endpoints, but the main application -- and the novelty compared with existing literature -- involves worldlines ending at the conformal boundary of AdS, at infinite proper distance. We could simply do all calculations with an infrared cutoff and take it to infinity as a final step, but as hinted above, things simplify if we take the cutoff away immediately.

When the separation $L$ between endpoints points is very large, the dominant contribution to the partition function grows exponentially in $L$, so $W_L = -\log Z$ has a piece proportional to $L$. From the point of view of perturbation theory, $W$ is the sum of connected diagrams, and for large $L$ the diagrams can be located anywhere along the particle trajectory; integrating over this location gives the linear growth. Roughly speaking, this is the ground-state energy as a function of the coupling\footnote{It is not quite the ground-state energy because of the massless mode $u$, in particular causing the choice of boundary conditions to be important. This zero mode is reflected in the $\log L$ piece of $W_L$ in \cref{twoLoopWL}, which could not appear in a quantum mechanics with discrete spectrum.}, which I denote by $\mathcal{E}$:
\begin{equation}\label{zeroE}
	\mathcal{E}(\lambda) = \lim_{L\to\infty} \frac{W_L(\lambda)}{L} \stackrel{\text{AdS}}{=} \frac{1}{2\lambda}+\frac{d}{2}+\frac{d^2}{8}\lambda
\end{equation}
The result quoted for the particle in AdS can be read off from the two-loop calculation \cref{twoLoopWL}, but in fact does not get corrected further at higher loops. This can be seen directly from explicit expressions of the heat kernel in hyperbolic space, for example in \cite{grigor1998heat}, taking the time and separation simultaneously to infinity.

This energy $\mathcal{E}$ will be altered when we include the effect of coupling to other fields, for example the gravitational field; these are just the usual self-energy corrections. On the other hand, modifying the theory in some finite region, for example by including its gravitational interactions with a second particle as we will do, will not alter $\mathcal{E}$, but only the finite pieces of $W_L$.

Writing the path integral in terms of the coupling $\lambda$, rather than the modulus $T$, we get
\begin{equation}
 \frac{L}{2}\int_0^\infty d\lambda\; e^{-\frac{m^2 \lambda}{2}L-W_L(\lambda)}
\end{equation}
which, at large $L$, will have a saddle point at the minimum of $\frac{m^2}{2}\lambda +\mathcal{E}(\lambda)$ as a function of $\lambda$, which will become exact in the large $L$ limit, even for finite $m$. The value of the exponent at the minimal $\lambda$ has a direct physical interpretation, the exponential fall-off of the two-point function giving the scaling dimension $\Delta$ of the dual CFT operator:
\begin{equation}\label{deltalambda}
	\Delta = \mathcal{E}(\lambda)-\lambda \mathcal{E}'(\lambda),\quad m^2 = -2 \mathcal{E}'(\lambda)
\end{equation}
This is simply a description of the mass renormalisation of the particle from the point of view of the worldline theory, and the `bare' mass $m$ (and coupling $\lambda$) do not have any independent physical meaning, and indeed will be cutoff dependent when we introduce interactions. However, for the free particle there is no such ambiguity, and substituting in the computed expression \cref{zeroE} for $\mathcal{E}(\lambda)$, we get $m^2 = \frac{1}{\lambda^2}-\frac{d^2}{4}$ and $\Delta = \frac{1}{\lambda}+\frac{d}{2}$, in particular reproducing the usual relation $m^2=\Delta(\Delta-d)$. Note in particular that negative $m^2$ does not present any particular problem beyond strong coupling, as long at is above the Breitenlohner-Freedman bound, but the dimension $\Delta$ must be above $\frac{d}{2}$, so it is not clear how to describe operators in the alternate quantisation regime $0<\Delta<\frac{d}{2}$ from this point of view.

The upshot of this is that we will never have to do an integral over the modulus, even in a saddle-point approximation as in the previous subsection at finite $L$. Instead, we compute the energy functional $\mathcal{E}$ (to the required order in perturbation theory) so we can express the coupling $\lambda$ in terms of the physical parameter $\Delta$, and subsequently perform all computations involving the interactions between particles at that fixed value of the coupling. Furthermore, we may take the boundary conditions to be fixed at infinity, with the parameter $s$ running over the whole real line (alternatively, if one end of the worldline ends at a vertex in the bulk and one end at the boundary, $s$ can run over $[0,\infty)$).

I conclude this subsection by writing down the Feynman rules for the AdS worldline theory with boundary conditions taken at infinity, in \cref{FeynmanRules2}.
\begin{table}
	\centering
	\begin{tabular}{m{3cm} m{4.5cm} m{4.5cm}}
	& \textbf{Position space} & \textbf{Frequency space} \\
		\begin{tikzpicture}
			\draw (0,0) -- (1,0) node [label=above:{$i$}]{} --(2,0);
			\path (0,-.6) -- (0,.6);
		\end{tikzpicture} & $\frac{\lambda}{2}e^{-|s_1-s_2|}$ & $\frac{\lambda}{1+\omega^2}$ \\
		\begin{tikzpicture}
			\draw[dashed] (0,0) -- (2,0);
			\path (0,-.6) -- (0,.6);
		\end{tikzpicture} & $-\frac{\lambda}{2}|s_1-s_2|  \qquad (\star)$ & $\frac{\lambda}{\omega^2} \qquad (\star)$\\
		\begin{tikzpicture}
			\draw [dashed]  (0,0) -- (-1,0) node [label=above:{$1$}]{};
			\draw (0,0) -- (.5, {sqrt(3)/2}) node [label=below:{$i$}]{};
			\draw (0,0) -- (.5, {-sqrt(3)/2}) node [label=above:{$j$}]{};
		\end{tikzpicture} & $-\frac{2}{\lambda}\delta_{ij} \partial_{s_1}$ & $\frac{2i}{\lambda}\omega_1\delta_{ij}$ \\
		\begin{tikzpicture}
			\draw [dashed]  (-.8,-.8) node [label=above:{$1$}]{} -- (0,0) -- (-.8,.8) node [label=below:{$2$}]{};
			\draw (0,0) -- (.8, .8) node [label=below:{$i$}]{};
			\draw (0,0) -- (.8, -.8) node [label=above:{$j$}]{};
		\end{tikzpicture} & $-\frac{2}{\lambda}\delta_{ij} \partial_{s_1}\partial_{s_2}$ & $\frac{2}{\lambda}\omega_1\omega_2\delta_{ij}$ \\
		\begin{tikzpicture}
			\draw (-.8,-.8) node [label=above:{$l$}]{} -- (0,0) -- (-.8,.8) node [label=below:{$i$}]{};
			\draw (0,0) -- (.8, .8) node [label=below:{$j$}]{};
			\draw (0,0) -- (.8, -.8) node [label=above:{$k$}]{};
		\end{tikzpicture} & $\frac{1}{\lambda}\left(\delta_{ij}\delta_{kl} \partial_{i}\partial_{l}+\text{11 perms}\right)$ & $-\frac{1}{\lambda}\left(\delta_{ij}\delta_{kl} \omega_{i}\omega_{l}+\text{11 perms}\right)$
	\end{tabular}
	\caption{\label{FeynmanRules2}The Feynman rules for the AdS worldline quantum mechanics with boundary conditions set at infinity, with $q$ propagators (solid line, indices $i,j,\ldots=1,2,\ldots,d$) and $u$ propagator (dashed line). As indicated by $(\star)$, the $u$ propagator is not strictly correct as the boundary conditions have not been properly enforced. The main effect is to eliminate the zero mode, so diagrams with $\omega=0$ running on $u$ propagators can be discarded. There are additional vertices similar to the final one for any even number of $q$ propagators.}
\end{table}
There is a minor subtlety with the $u$ propagator, since we are putting boundary conditions at infinity for a particle of zero mass. This does not cause too much difficulty, because the $u$ propagator is always differentiated at both ends, though the boundary conditions have the important effect of removing the zero mode. This means that with these boundary conditions, we can drop diagrams with zero momentum on a $u$ propagator, such as the two-loop vacuum graph \begin{tikzpicture}[baseline={([yshift=-.5ex]current bounding box.center)}]
			\node[draw,circle,minimum size=15](A) at (0,0){};
			\node[draw,circle,minimum size=15](B) at (1,0){};
			\draw[densely dashed] (A) --(B);
		\end{tikzpicture}. Formally, if the propagator is $G_u(s_1,s_2)$ we can use $\partial_1\partial_2G_u(s_1,s_2)=\lambda(\delta(s_1-s_2)-L^{-1})$. If in any doubt, one can always go back to the rules in \cref{HKQM} with boundary conditions at finite times, where there is no ambiguity, and take $L$ to infinity at the end.

\subsection{Vertex operators and their correlation functions}\label{VOcorrelators}

With this understanding of the free particle under our belt, we can finally start to explore what happens when we allow it to interact. From the point of view of the worldline, the interactions are included by coupling to external fields, though we will subsequently integrate over those fields. I will focus here on the coupling to gravity, which means allowing the metric to fluctuate. Writing the metric as a perturbation around pure AdS, $g=g_{AdS}+h$, this means we simply deform the AdS theory by inserting the operator $\exp\left(-\frac{1}{2\lambda}\int ds h_{ab}(x)\dot{x}^a\dot{x}^b\right)$, and if we expand for small fluctuations, the exponential can be expanded. The result is that at each order we can simply insert a number of \emph{vertex operators} in the path integral:
\begin{equation}\label{VO}
	\mathcal{V}_h = \frac{1}{2\lambda}\int ds \;h_{ab}(x(s))\dot{x}^a(s)\dot{x}^b(s)
\end{equation}
This vertex operator is not in fact quite correct, since the measure must also be modified; this can be achieved by adding insertions of the ghosts discussed towards the end of \cref{WLpert}, but we will simply correct it by hand when required later. The inclusion of vertex operators is much the same as in string theory, where modifying the background is achieved by inserting marginal scalar operators integrated over the worldsheet. Often, working around flat space in particular, the background field configuration is written as a plane wave of definite momentum and polarisation ($h_{ab}(x)=\epsilon_{ab}e^{ip\cdot x}$), but I will keep it as an arbitrary function.

Now we simply need to compute correlation functions of vertex operators! In full generality, this is a formidable task, but luckily we have already developed the necessary machinery to use perturbation theory in the coupling $\lambda\approx\frac{1}{m}$. For this, we merely need to write $x^a$ in terms of our perturbations $u$ and $q^i$, Taylor expand to whatever order we desire, and finally compute correlation functions of the monomials in $q$ and $u$ appearing, helped by $O(d)$ and time-reversal symmetries. Expanding to second order (and integrating by parts whenever $h$ is differentiated in the $t$ direction) the vertex operator is
\begin{align}
	\mathcal{V}_h=&\frac{1}{2\lambda}\int ds\Big[h_{00}+(\dot{u}+q^i\partial_i)h_{00}+2\dot{q}^i h_{0i}\\&+\left(\frac{1}{2}q^iq^j\partial_i\partial_j+\dot{u}q^i\partial_i-u\dot{q}^i\partial_i-u\ddot{u}\right) h_{00} +2\left(\dot{q}^iq^j\partial_j-u\ddot{q}^i\right)h_{0i}+\dot{q}^i\dot{q}^j h_{ij}+ \cdots\Big]\nonumber
\end{align}
plus ghosts added for good measure. % guffaw
 Here the metric perturbation and its derivatives are evaluated at $t=s,q=0$, corresponding to the unperturbed classical solution.

For the one-point function, most of the terms vanish simply by using $O(d)$ and time-reversal invariance, with the nonzero expectation values appearing at this order being the following:
\begin{equation}
	\langle q^i q^j \rangle = \frac{\lambda}{2}\delta^{ij},\quad \langle u \ddot{u}\rangle = -\lambda\delta(0),\quad \langle \dot{q}^i \dot{q}^j \rangle = \lambda\left(\delta(0)-\frac{1}{2}\right)\delta^{ij}
\end{equation}
The $\delta(0)$ divergences appear here from differentiating factors like $|s-s'|$ twice and setting $s=s'$, and when the path integral is regulated by discretisation, $\delta(0)$ is simply the inverse of the step size in $s$. These contributions are cancelled by proper inclusion of the measure factor, which can be accounted for at leading order in the metric perturbation by directly including the factors of the determinant of the metric as additional insertions in the path integral:
\begin{align*}
	\prod_s\sqrt{\det(g(s)+h(s))} &= \exp\left(\frac{1}{2}\sum_s \log \det(g(s)+h(s))\right) \\ &=1+\frac{\delta(0)}{2}\int ds \; g^{ab}(x(s))h_{ab}(x(s))+O(h^2)
\end{align*}
In the last line I have expanded the determinant and exponential to linear order in $h$ by Jacobi's formula, used the fact that $\det g=1$ here, and finally replaced the sum with an integral, including the factor of $\delta(0)$ to account for the step size in the discretised path integral. The final result for the one-point function of $\mathcal{V}_h$ is that we should subtract
\begin{equation}
	\frac{\delta(0)}{2}\int ds \left\langle \frac{h_{00}(x(s))}{1+q^2}+(\delta^{ij}+q^iq^j)h_{ij}(x(s))\right\rangle = \frac{\delta(0)}{2}\int ds\; \left(h_{00}+h_{ii}\right) +O(\lambda)
\end{equation}
from the na\"ive result, which acts precisely to cancel the troublesome delta-functions.

Putting this all together, we get the one-point function of the vertex operator to one-loop order:
\begin{equation}\label{V1loop}
	\langle\mathcal{V}_h\rangle = \int ds\;\left( \frac{1}{2\lambda}h_{00}+\frac{1}{8}\partial_{i}\partial_{i} h_{00}-\frac{1}{4}h_{ii}+O(\lambda)\right)
\end{equation}
This answer should be diffeomorphism invariant, so that if we choose $h$ to be pure gauge, the integrand is a total derivative of $s$. Classically, this is because we are perturbing a geodesic: a small diffeomorphism is equivalent to a first order perturbation of the particle path, which has vanishing variation on-shell. Indeed, it is straightforward to check that the particular combinations of terms appearing in the tree-level and one-loop contributions here vanish when $h$ is a Lie derivative of the background metric.

Similarly, we can compute the two-point function of the vertex operator $\langle\mathcal{V}_h\mathcal{V}_h\rangle$. This is dominated by the disconnected component, that is the square of the one-point function, so I record here the first nontrivial (tree-level) contribution to the connected correlator $\langle\mathcal{V}_h\mathcal{V}_h\rangle_c=\langle\mathcal{V}_h\mathcal{V}_h\rangle-\langle\mathcal{V}_h\rangle\langle\mathcal{V}_h\rangle$. This follows straightforwardly as before (in particular, the measure being irrelevant at this order), though the answer is more complicated since there is nontrivial dependence on the difference in times $s,s'$ of the two insertions of the metric perturbation.
\begin{align}\label{VO2pt}
	\langle\mathcal{V}_h&\mathcal{V}_h\rangle_c^{\text{(tree)}} =\frac{1}{4\lambda^2}\int ds\,ds'\,\left\langle \left((\dot{u}+q^i\partial_i)h_{00}+2\dot{q}^i h_{0i}\right)(s)\left((\dot{u}+q^i\partial_i)h_{00}+2\dot{q}^i h_{0i}\right)(s')\right\rangle \\
	&=\frac{1}{4\lambda}\int ds\; h_{00}^2%+4 h_{0i}h_{0i}\right)
	+\frac{1}{8\lambda}\int ds\,ds'\;e^{-|s-s'|}\left(\partial_ih_{00}(s)-2\dot{h}_{0i}(s)\right)\left(\partial_ih_{00}(s')-2\dot{h}_{0i}(s')\right) \nonumber
	%\\+\frac{1}{8\lambda}\int ds\,ds'\;&e^{-|s-s'|}\left[\partial_i h_{00}(s)\partial_i h_{00}(s')+4\operatorname{sgn}(s-s')\partial_i h_{00}(s) h_{0i}(s')-4h_{0i}(s)h_{0i}(s') \right]\nonumber
\end{align}

It is clear how to proceed further, computing higher point correlation functions, going to larger loop order, or considering coupling to other fields, for example gauge fields. We could also describe particles with spin, with perhaps the most interesting direction to develop being the description of fermions, which is achieved by promoting the worldline theory to a supersymmetric quantum mechanics \cite{Brink:1976sz,Brink:1976uf}. But we have plenty here to start attacking the problem of full correlation functions in AdS, which requires treating the metric not simply as an external field, as we have so far, but making it dynamical.

\section{Four-point functions\label{4pt}}

Now that we know how to compute vertex operator correlation functions, we can start calculating the diagrams described in \cref{diagrammatica}, to find contributions to the four-point function $\langle\phi_1\phi_1\phi_2\phi_2\rangle$ of heavy operators $\phi_{1,2}$, of conformal dimensions $\Delta_1,\Delta_2$. Here, I have slightly generalised to allow for two different species of particle, since this does not add any more difficulties, so we have one worldline corresponding to particle $1$ connecting the insertion points of $\phi_1$ on the conformal boundary, and similarly one for particle $2$.

Starting with two of geodesics in AdS, one for each particle, the four-point function is given by the sum over all vacuum diagrams, which is the exponential of the sum of connected diagrams. We can simplify things slightly more by dividing out by the trivial connected correlator $\langle\phi_1\phi_1\rangle\langle\phi_2\phi_2\rangle$, which simply removes diagrams that interact only with one worldline or the other, leaving only connected diagrams that also connect the two particles.

The diagrams interacting with only one of the particles still have a r\^ole to play, determining the two-point functions, but because of the symmetry this is encoded in a single parameter, namely the dimension $\Delta$. But the interaction between the particle and gravity will renormalise the dimension, by contributing to $\mathcal{E}$ as described around \cref{deltalambda}; these diagrams therefore will correct the value of the worldline coupling $\lambda$ at which we must compute. Furthermore, these contributions due to the interaction with bulk quantum fields will be divergent, so we must regularise, and the renormalisations of $\lambda$ will be formally infinite. This fixing of the value of $\lambda$ in terms of $\Delta$ and bulk couplings is the only thing that remains of the integral over the modulus $T$ discussed at length in the previous section.

\subsection{Conformal blocks and geodesic Witten diagrams}

The leading order diagram contributing to the connected correlator describes a single graviton being exchanged between the worldlines, coupling by the tree-level vertex operators at either end.
\begin{equation}
	\begin{tikzpicture}[baseline={([yshift=-.5ex]current bounding box.center)},scale=0.6, every node/.style={scale=0.6}]
%		\draw [help lines] (0,0) grid (8,8);
		\draw [thick] (2,2) circle(2);
		\draw [dashed] (.6,.6) arc (-45:45:2);
		\draw [dashed] (3.4,3.4) arc (135:225:2);
		\node [draw,circle,cross,inner sep=0, minimum size=5](A) at (1.1,1.4){};
		\node [draw,circle,cross,inner sep=0, minimum size=5](B) at (2.88,2.5){};
		\draw [grav]  (A) -- (B);
	\end{tikzpicture}= \frac{1}{2}\langle \langle\mathcal{V}^{(1)}_h\rangle^{(1)}_\text{tree} \langle\mathcal{V}^{(2)}_h\rangle^{(2)}_\text{tree} \rangle_\text{gravity} = \frac{\Delta_1\Delta_2}{8} \int ds_1\,ds_2 \langle h_{00}(s_1) h_{00}(s_2) \rangle
\end{equation}
The rather clumsy notation for the correlation function is designed to spell out exactly what is being computed in great detail, with $\mathcal{V}^{(1)}_h$ denoting the vertex operator for $h$ associated to worldline $1$, $\langle\cdot\rangle^{(1)}_\text{tree}$ denoting the tree-level (order $\Delta$) correlation function in the quantum mechanics living on that worldline, and $\langle\cdot\rangle_\text{gravity}$ denoting the expectation value of the metric fluctuations in the usual bulk Einstein-Hilbert theory. These details will subsequently be dropped. The $\frac{1}{2}$ is a symmetry factor. By $h_{00}$ I mean the metric fluctuation dotted into the tangent vector of the corresponding geodesic.

It only remains to evaluate the integral appearing on the right hand side, which is the bulk-to-bulk graviton propagator between two geodesics, dotted into their tangent vectors, and integrated over their lengths. But this precise calculation has been done before! It is the definition of a `geodesic Witten diagram' for the graviton \cite{1508.00501}. In fact, they include bulk-to-boundary propagators for the external fields, but, at least when the external operators are identical in pairs as here, these simply give the overall factor of the disconnected correlator and do not affect the integral. The original motivation for introducing this object was to find a bulk dual of a conformal block, the natural kinematic objects of the dual CFT, packaging the contribution to the correlator from exchanging an operator and its descendants. This diagram is therefore simply proportional to the conformal block $\mathcal{F}_T$ giving stress-tensor exchange. Choosing a different internal field, rather than the graviton, would give the analogous answer, so quite generally we can reproduce a conformal block in this way. Restricting the integral to the geodesic is not a choice here, but follows inevitable from the worldline approach.

Remaining %remoaner
 with one-graviton exchange diagrams, we can consider computing to higher loops in the vertex operator correlation functions, but the high degree of symmetry means that the higher loops do not change the answer much. This follows simply from $O(d)$ and time-reversal symmetry of the worldline theory: the vertex operator correlation functions can only depend on $h_{00}$, the spatial trace $h_{ii}$, the contraction $\partial_i\partial_jh_{ij}$, and spatial Laplacians $\partial_i\partial_i$ acting on these any number of times. Picking some gauge, for example transverse-traceless ($h_{ab}g^{ab}=0,\nabla^a h_{ab}=0$), and using the free linearised equations of motion to eliminate the spatial Laplacians, we can eliminate all terms in favour of $h_{00}$ (for an explicit example doing this at one loop, see the next paragraph). The free equations of motion are relevant since the propagator obeys them away from the source, and as long as the two geodesics do not intersect, the source never contributes. This works similarly for any choice of exchanged field. To any loop order, the answer is therefore still proportional to the conformal block, with only the constant of proportionality being altered. These higher-loop corrections to the vertex operator one-point function fix the normalisation of the geodesic Witten diagram in terms of the worldline parameters.

For most bulk fields, the couplings of the vertex operator to the worldline theory are free parameters, so this normalisation just determines the physical three-point function in terms of the bare parameters. For gravity, however, the couplings are constrained by diffeomorphism invariance, and the three-point function $\langle\phi\phi T\rangle$ is determined by Ward identities in terms of the dimension $\Delta_\phi$, so matching the OPE coefficient with the dimension \cref{deltalambda} determined by the long-distance behaviour of the propagator provides a nontrivial check. At leading order in $G_N$, without graviton interactions (but to all orders in $\lambda$), this means that the vertex operator one-point function should be proportional to $\Delta$, at least when the metric perturbation $h$ is on-shell, since the graviton propagator satisfies the linearised Einstein equations. The one-loop contribution to $\langle \mathcal{V}_h\rangle$ in \cref{V1loop} is $\frac{1}{8}\int ds\,(\partial_{i}\partial_{i} h_{00}-2 h_{ii})$, which we would like to rewrite as something proportional to $\int h_{00}$ using the equations of motion for the linearised metric as described in the previous paragraph. Choosing transverse-traceless gauge $h^a_a=0$, $\nabla^ah_{ab}=0$ (we have already verified gauge invariance of this one-loop expression), the equation of motion is $(\nabla^2+2)h_{ab}=0$, from which we need the $00$ component evaluated at $q=0$, which gives $(\partial_0^2+\partial_i\partial_i-2d)h_{00}=0$. Eliminating $\partial_{i}\partial_{i} h_{00}$ with this, dropping the $\partial_0$ derivative, and using tracelessness $h_{00}+h_{ii}=0$, we get the on-shell transverse-traceless one-loop vertex operator one-point function % ninety-seven-horsepower omnibus
 $\langle \mathcal{V}_h\rangle = \left(\frac{1}{2\lambda}+\frac{d}{4}+O(\lambda)\right)\int ds\; h_{00}$. Finally we can substitute the coupling in favour of the dimension using $\lambda=\frac{1}{\Delta-\frac{d}{2}}$ from \cref{deltalambda}, and find a result proportional to $\Delta$ as expected, the one-loop renormalisation of the coupling $\lambda$ exactly compensating for the one-loop contribution to the vertex operator one-point function. Going to higher loops, the formula given for $\lambda$ in terms of $\Delta$ receives no further corrections, so the higher-loop contributions to $\langle \mathcal{V}_h\rangle$ should all vanish when $h$ is on-shell. There will be further corrections to this at next order in $G_N$, with interactions of the graviton in the computation of the correlation functions, and including the (divergent) gravitational self-energy diagram in the computation of $\mathcal{E}(\lambda)$, which renormalises $\lambda$ at order $G_N$.

From this perspective, the geodesic Witten diagram does not truly localise on the worldline, since loop corrections to the vertex operator expectation value depend on higher and higher derivatives of the metric, and it is only the high symmetry of the situation that makes these corrections trivial. Strictly localising to the worldline does not work even in thermal AdS for example \cite{1706.00047}, discussed in \cref{thermal}, where the loops give nontrivial corrections. Apart from the symmetry, the other crucial aspect that makes the geodesic Witten diagram work is that the source term in the equation of motion acting on the propagator does not contribute. From the current point of view, this is a result of considering the worldline quantum mechanics only in perturbation theory; we will see shortly how this is modified by non-perturbative corrections, which provide the additional double-trace contribution to the Witten diagram.

This phenomenon of isolating conformal blocks of single-trace exchanges is in fact rather general, discussed more in \cref{thermal}. It is another virtue of describing bulk physics in worldline language, in that it makes the connection to boundary conformal field theory language rather closer and more natural. In retrospect, this should be unsurprising, since the worldline, like the standard CFT language, separates states by particle number, while a quantum field packages all single- and multi-particle states together.

A useful new perspective on geodesic Witten diagrams was introduced in \cite{1604.03110}, which broke the diagram up further into a correlation function of nonlocal operators dubbed `OPE blocks'. It is natural to identify these OPE blocks with the one-point functions of vertex operators $\langle\mathcal{V}_h\rangle$. This is explored a little more in \cref{OPEblocks}.

\subsection{Exponentiation}

Understanding the appearance of the exchange conformal block as this diagram immediately implies something interesting, because we can also include multiple disconnected copies of the same diagram. This means that in the full 4-point function, the exchange conformal block exponentiates!
\begin{equation}
	\frac{\langle \phi_1\phi_1\phi_2\phi_2\rangle}{\langle\phi_1\phi_1\rangle\langle\phi_2\phi_2\rangle} = e^{\Delta_1\Delta_2G_N \mathcal{F}_T + O(\Delta^3 G_N^2)}
\end{equation}
For example, if we take a limit where $\Delta_1,\Delta_2$ go to infinity and $G_N$ to zero, but with $\Delta_1\Delta_2 G_N$ held fixed, the correlation function approaches the exponential of the block. Note that this involves summing contributions from Witten diagrams at arbitrarily high loops. This phenomenon has already observed for gravity in $d=2$, where these graviton exchanges are included in a conformal block for the extended Virasoro symmetry (of which more later) \cite{1403.6829,1504.01737}. The exponential of the block (the function $\mathcal{F}_T(x)=x^2 {}_2F_1(2,2,4,x)=6 \left(\left(1-\frac{2}{x}\right) \log (1-x)-2\right)$ in this case) appears by directly summing the relevant descendant contributions using the Virasoro algebra\footnote{The authors of these papers apparently found this surprising enough to warrant an exclamation mark!}. It also makes an appearance for large spin operators in a certain kinematic limit \cite{1504.01737}, to solve crossing symmetry. But this analysis shows it must be a much more general phenomenon, applying to holographic theories in general dimension, and with general fields exchanged, since the same calculation holds if a scalar, gauge or other field mediates the interaction.

It would be interesting to show this exponentiation directly from bootstrap arguments, using crossing symmetry and assumptions of a large gap, generalising the arguments of \cite{1504.01737}. Conversely, one could assume exponentiation, and use this to extract data about OPE coefficients of multi-trace operators from those with smaller numbers of traces. For example, the order $\Delta^2 G_N$ part of the OPE $\phi\phi[TT]_{0,0}$ of $\phi$ with the lightest stress-tensor double trace can be read off immediately from expanding the exponential to order $x^4$ in the cross-ratio, and something similar continues for higher multi-traces. The exponentiation continues to higher loops, and higher order in $G_N$, providing a potentially rich set of constraints on multi-trace operators.

\subsection{Higher-order contributions}

\begin{figure}
	\centering
	\begin{subfigure}[b]{.45\textwidth}
	\begin{tikzpicture}[scale=0.8, every node/.style={scale=0.8}]
%		\draw [help lines] (0,0) grid (8,8);
		\draw [ultra thick] (4,4) circle(4);
		\draw [thick, dotted] (1.2,1.2) arc (-45:45:4);
		\draw [thick, dotted] (6.8,6.8) arc (135:225:4);
%		\draw [thick] (2.3,4.4) circle(.3);
%		\draw [thick] (5.75,3.1) circle(.3);
%		\draw [grav] (2.57,4.3) -- (5.45,3.2);
		\node [draw,thick,circle,cross,inner sep=0, minimum size=7](A) at (2.2,5.2){};
		\node [draw,thick,circle,cross,inner sep=0, minimum size=7](D) at (2.2,2.8){};
		\node [draw,thick,circle,fill=black,inner sep=0pt,minimum size=4](B) at (5.7,4.7){};
		\node [draw,thick,circle,fill=black,inner sep=0pt,minimum size=4](C) at (5.7,3.3){};
		\draw [grav] (A) -- (B);
		\draw [grav] (C) -- (D);
		\draw [thick] (B) --(C);
	\end{tikzpicture}
	\caption{\label{2graviton1}}
	\end{subfigure}
	\begin{subfigure}[b]{.45\textwidth}
	\begin{tikzpicture}[scale=0.8, every node/.style={scale=0.8}]
%		\draw [help lines] (0,0) grid (8,8);
		\draw [ultra thick] (4,4) circle(4);
		\draw [thick, dotted] (1.2,1.2) arc (-45:45:4);
		\draw [thick, dotted] (6.8,6.8) arc (135:225:4);
%		\draw [thick] (2.3,4.4) circle(.3);
%		\draw [thick] (5.75,3.1) circle(.3);
%		\draw [grav] (2.57,4.3) -- (5.45,3.2);
		\node [draw,thick,circle,cross,inner sep=0, minimum size=7](A) at (5.63,3.7){};
		\node [draw,thick,circle,cross,inner sep=0, minimum size=7](B) at (2.2,2.8){};
		\node [draw,thick,circle,cross,inner sep=0, minimum size=7](C) at (2.3,4.6){};
		\coordinate (int) at (4,3.7);
		\draw [grav] (C) -- (int);
		\draw [grav] (int) -- (B);
		\draw [grav] (int) -- (A);
	\end{tikzpicture}
	\caption{\mercury\label{mercury}}
	\end{subfigure}
	\caption{The two diagrams appearing at tree-level, two-graviton order, $m^3G_N^2$.\label{2gravitons}}
\end{figure}
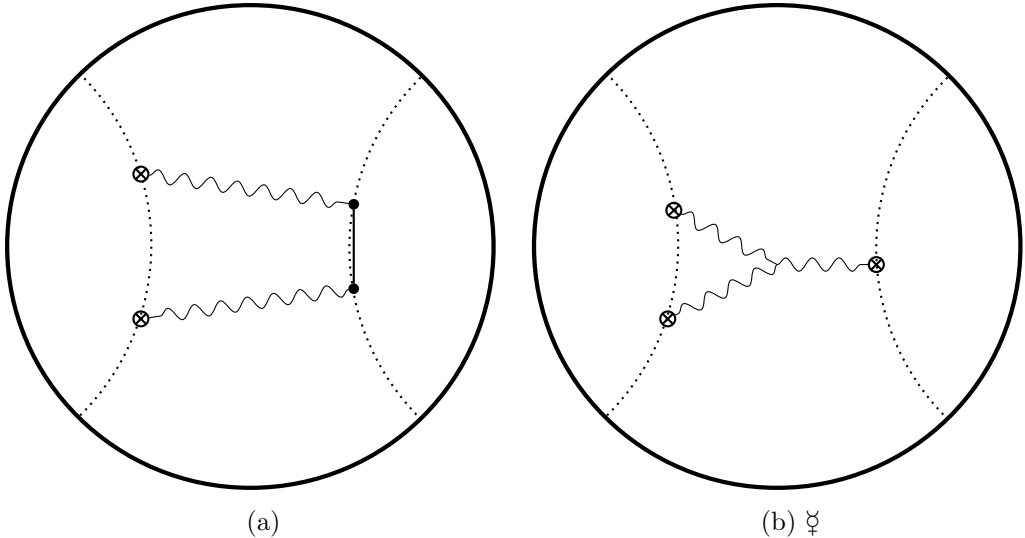

Having accounted for all the diagrams of order $G_N$, with a single graviton exchanged, we now move to the next order in perturbation theory by adding an extra graviton, with two diagrams contributing at  $O(m_1^2 m_2 G_N^2)$, shown in \cref{2gravitons}. These diagrams are tree-level, capturing classical physics when formulated in terms of particles, though from the point of view of fields and Witten diagrams, they first appear at one loop, as contributions to box and penguin diagrams (and, as before, reappear at higher loops through exponentiation).

The first diagram, in \cref{2graviton1}, accounts for the second order change in the worldline length of particle 2 due to the linearised gravitational field created by particle 1. Roughly speaking, one of the vertices on the second worldline perturbs the metric, the propagator encodes the effect of that perturbation on the location of the geodesic, and the second vertex computes how that displacement of the geodesic changes the length. The graviton propagators calculate the metric perturbation sourced by worldline 1.

Very explicitly, this second order perturbation of the geodesic length is encoded in the tree-level vertex operator connected two-point function
$\langle\mathcal{V}_h\mathcal{V}_h\rangle_c^{\text{(tree)}}$, calculated in \cref{VO2pt}. One alternative way to compute this formula is by directly solving for the location of a geodesic in a perturbed background to first order, with the result
\begin{align}
	q^i(s) &= \frac{e^s}{2}\int_s^\infty ds'\; e^{-s'} \left(\dot{h}_{0i}(s')-\frac{1}{2}\partial_i h_{00}(s')\right) + \frac{e^{-s}}{2}\int_{-\infty}^s ds'\; e^{s'} \left(\dot{h}_{0i}(s')-\frac{1}{2}\partial_i h_{00}(s')\right) \nonumber \\
	\dot{u}(s) &= -\frac{1}{2}h_{00}(s),
\end{align}
and then substituting this back in the integral expression for the length. This matches \cref{VO2pt}, with the only difference being a symmetry factor of 2. All that remains is to substitute in the metric perturbation $h$, which is sourced by the other worldline\footnote{This is precisely the same as the object $h_{\nu_1,\ldots,\nu_\ell}$ in equation (5.6) of \cite{1508.00501}.}:
\begin{equation}\label{linearfield}
	h_{ab}(x) = m_1 G_N \int_\text{WL1} ds \; G_{ab,cd}(x,x_1(s))\;\dot{x}_1^c(s)\dot{x}_1^d(s)
\end{equation}
Here, $G_{ab,cd}(x,y)$ denotes the graviton propagator, and $x_1(s)$ is the geodesic of particle 1, so we are taking the component of the propagator in the direction of its tangent vector. It is straightforward to explicitly compute this linearised metric, as done for example in \cite{1508.00501}, and, in the end, the result of the full diagram \ref{2graviton1} is the second order correction to the action of worldline 2 in the linearised background created by worldline 1 as claimed (though I will not write down explicit expressions for the result). Using this approach of computing the metric sourced by the worldline is advantageous when possible, avoiding use of the complicated explicit expressions for the graviton propagator \cite{hep-th/9902042}.

The second diagram appearing at the same order, \cref{mercury}, computes the leading order correction to the action of worldline 2, but now going to second order in the background perturbation, giving the first nonlinear correction to the gravitational field sourced by worldline 1 at order $(m_1G_N)^2$. This diagram is the AdS analogue of Einstein's famous computation of the precession of the perihelion of Mercury \cite{einstein1914erklarung}. Direct computation of the diagram is rather involved, but ultimately is equivalent to solving Einstein's equations to quadratic order with a static, spherically symmetric ansatz. Computing the analogue of \cref{linearfield} with the second-order metric fluctuation, the equation of motion is now sourced by the square of the linearised perturbation, with the appropriate derivatives and index contractions, given Einstein's equations to quadratic order. The only subtlety to deal with is the contact term (and potential divergence) from when all vertices come together, related to the renormalisation of the mass. It would be instructive to work this out in detail, and useful for doing higher-order calculations.

Going to higher orders in $m_1G_N$, including all diagrams of order $(m_1 G_N)^n m_2$, the pattern continues, including higher-order corrections to the metric sourced by particle $1$ coming from graviton interactions, as well as corrections to the geodesic along which particle $2$ moves in the shifted background from higher-point (but still tree-level) vertex operator correlation functions. The result is the length of a geodesic corresponding to particle 2 in the Schwarzschild-AdS metric sourced by particle 1, computed perturbatively in $m_1G_N$.

Put in another way, the summation of diagrams to all orders in $m_1 G_N$ solves the full nonlinear classical equations of motion for particle 1 coupled to gravity, while still treating particle 2 in the probe limit without backreaction. Since we have an explicit description of the background sourced by one particle, it would be nice to use this as the starting point for perturbation theory in $m_2$ and $G_N$, but because of the singular nature of the Schwarzschild metric it is not clear how to keep control of the worldline theory of particle 1. The notable exception to this is $d=2$, where for sufficiently small (but finite) $m_1G_N$ the metric only has a mild conical singularity; this will be discussed in \cref{VirDisc}.

Staying at tree level, we can also include corrections in powers of $m_2 G_N$, accounting for the backreaction of particle 2 as well as particle 1. This provides a systematic expansion computing the metric sourced by a pair of interacting massive particles order by order in the strength of the gravitational field. This is closely related to the post-Newtonian expansion, used \emph{inter alia} to compute the gravitational wave radiation from the inspiral phase of black hole mergers \cite{Abbott:2016blz}. This expansion can be cast similarly in the form of worldline diagrams \cite{hep-th/0409156}, though the expansion parameters are slightly different.

Finally, we can also introduce genuine quantum corrections to the amplitude in powers of $m^{-1}$, starting with one-loop diagrams at order $m_1m_2G_N^2$, $m_1^2G_N^2$ and $m_2^2G_N^2$. These calculations cannot be done with such minimal effort as expended so far, but they are more tractable than computing loop Witten diagrams. I leave this technical development, and exploration of the results, to future study.

\subsection{The Virasoro blocks}

We are describing some object order by order in perturbation theory, which in the free limit $G_N\to 0$ at fixed dimension approaches the global conformal block $G_N \Delta^2 \mathcal{F}_T$, so a natural characterisation of this object is a gravitationally corrected, or interacting, conformal block. In most contexts this is something new, but in the special case $d=2$, something familiar already fits this description, namely the Virasoro conformal block. I claim that in this case, they are in fact the same, my construction providing a bulk definition of the vacuum Virasoro block\footnote{By this I mean the scalar Virasoro block, which is the product of holomorphic and antiholomorphic blocks with equal dimensions $h=\bar{h}=\frac{\Delta}{2}$. A nice property of the worldline description is that there is an obvious generalisation to operators with spin by altering the worldline quantum mechanics \cite{Brink:1976sz,Brink:1976uf,hep-th/0101036}, which should reproduce blocks with $h\neq \bar{h}$.}.

The justification of this is relatively straightforward when we understand what physics the Virasoro block includes and, more importantly, what it excludes. It packages the contribution to a correlation function from the identity and its Virasoro descendants, namely the stress tensor and its derivatives, and all the multi-trace stress tensor operators. Multi-trace operators are not defined in a generic CFT, usually only existing in a perturbative large-$N$ sense, but in $d=2$ the stress-tensor multi-traces are defined in terms of the extended symmetry algebra. These operators are dual to multi-particle graviton states. Importantly, however, the full correlation function must include other operators to satisfy crossing symmetry. In our context of a large $c$ theory with a gap, described in the the bulk by our scalars weakly coupled to gravity, the extra exchanges are provided by the double-trace operators constructed from our external scalars \cite{0907.0151}, and multi-traces at higher order in the large $c$ expansion, all operators with dimensions that scale with $\Delta$. The Virasoro block therefore includes all the exchanged operators whose dimension stays finite when we take both $\Delta$ and $c$ to be large, which is precisely the contribution to the correlation function captured by a perturbative expansion in $\frac{\Delta}{c}$ and $\frac{1}{\Delta}$: exchanged operators of large dimension $\Delta_p\gg 1$ are exponentially suppressed relative to the free correlator by the exchange dimension $\Delta_p$, by the factor $x^{\Delta_p}$. In summary, \emph{the Virasoro vacuum block is the result of correcting the factorised correlation function to all orders in perturbation theory in $\frac{\Delta}{c}$ and $\frac{1}{\Delta}$}. This is precisely the perturbative expansion described in bulk language by worldlines coupled to gravity, so this must be a bulk description of the same object. I emphasise that this is \emph{not} the same as the sum of Witten diagrams, so the vacuum Virasoro block does not include all the gravitational contributions to a correlation function, and despite occasional claims to the contrary, gravitational physics in AdS$_3$ is not determined by kinematics alone.

This suppression of double-trace exchanges is an important part of the philosophy of `vacuum block domination', pioneered in \cite{1303.6955}, the idea that the correlation function in the semiclassical limit is given by the Virasoro vacuum block, in whichever channel it is largest. The double-traces nonetheless contain interesting physics, such as the gravitational binding energy of particles.

A reasonable complaint at this point is that I have attempted to define the Virasoro block as the summation of a perturbation theory, but this series cannot possibly converge (there are poles in the block at null states, which accumulate at $c=\infty$ and $\Delta=\infty$). One simple way to evade this is to compute order-by-order in a power series expansion in cross-ratio $x$ or nome $q$, in which case the coefficients are rational functions of $\Delta$ and $c$, with finite radius of convergence. Perhaps better than this, the blocks can be expressed as a convergent sum over global blocks \cite{1502.07742}, and the perturbative worldline expansion is well-suited to perturbatively compute the coefficients appearing in this expansion, again being rational functions of $c$ and $\Delta$. The coefficients to expand any given diagram in terms of global conformal blocks are determined uniquely, for example using the inversion formula \cite{1703.00278}.

The discussion here is perhaps a little abstract, and it helps to have specific examples in mind where this reproduces known results in the literature. The Virasoro blocks have been the subject of a lot of study, particularly in various limits of large $c$ and large dimensions as is most relevant here, and \cref{VirDisc} contains a detailed discussion of how the worldline formalism connects to this existing literature, which includes various consistency checks of the claimed equivalence.

\subsection{Double-trace saddle point \label{DT}}

I complete the section by discussing contributions to the correlation function that are not included in the perturbative expansion around the factorised correlator. As discussed previously, the full worldline path integral should result in the propagator, to reproduce the usual Feynman rules. For example, if we go to first order in $G_N$ only, but compute exactly on the worldline, this includes one graviton exchanged between the particles and should result in the exchange Witten diagram (first computed in \cite{hep-th/9903196}). But we have seen that, to all orders in worldline perturbation theory, the result is the exchange conformal block, and the Witten diagram is known to have additional contributions corresponding to the conformal blocks of double trace operators $[\phi_1\phi_1]_{n,l}$ and $[\phi_2\phi_2]_{n,l}$, describing two-particle states of the external operators (see \cite{1706.00047} for a review, and references therein). To account for this discrepancy, note that exchanged operators of dimension $\Delta_p$ contribute to the correlation function with a kinematic factor $x^{\Delta_p}$, and the double-trace contributions have dimensions $\Delta_p$ of order the external dimensions $2\Delta_{1,2}$ or larger. Therefore (unless the OPE coefficients scale exponentially in $\Delta$ to compensate) the double traces are exponentially suppressed in $\Delta$, and so are non-perturbative in the $\Delta^{-1}$ expansion. Nonetheless, I propose that the double-traces can be captured in the worldline formalism through a nonperturbative `instanton' contribution, an extra saddle-point in the worldline path integral. I leave full investigation of this to the future, just giving some preliminary hints for how this might work.

For this section, to simplify the exposition I will work not with a graviton exchange, but a scalar dual to an operator $\phi_p$ of dimension $\Delta_p$, with simple (non-derivative) cubic interaction vertices with $\phi_1\phi_1$ and $\phi_2\phi_2$. The extra saddle-point is slightly subtle, since it is not visible in the strict probe geodesic limit, but perhaps the easiest way to see it is directly from the integrand of the exchange Witten diagram, with bulk vertices at $x,y$:
\begin{equation}\label{networkAction}
	G^{(1)}_{b\partial}(x_1,x)G^{(1)}_{b\partial}(x_2,x)G^{(p)}_{bb}(x,y)G^{(2)}_{b\partial}(x_3,y)G^{(2)}_{b\partial}(x_4,y)=e^{-S^{(1)}(x)-S^{(2)}(y)-\Gamma^{(p)}(x,y)}
\end{equation}
In this expression, I have written the bulk-to-boundary propagators in terms of the classical actions for the worldlines, so $S^{(1)}(x)$ is $\Delta_1$ times the sum of regulated lengths from $x$ to the boundary points $x_1,x_2$, and similarly for $S^{(2)}(x)$. I have written the bulk-to-bulk propagator in terms of $\Gamma^{(p)}(x,y)=-\log G^{(p)}_{bb}(x,y)$, which, using the worldline formalism for the exchanged field as well as the external fields, can be understood as the worldline effective action for the internal particle, including loop effects. These quantum corrections are required (even if we also take $\Delta_p$ to be large) because, as we will see, the short-distance singularity of the propagator is important.

A nice intuition is to think of $S^{(1)}(x)+S^{(2)}(y)+\Gamma^{(p)}(x,y)$ as the energy of a network of elastic strings, with vertices at $x$ and $y$, and tensions given by the derivative of the relevant term with respect to the length of the string. A saddle-point in the Witten diagram integral is an extremum of this energy, corresponding to an equilibrium of the string network. The external operator worldlines have constant tensions $\Delta_1,\Delta_2$, which we are taking to be large, so under most circumstances the internal operator has an insignificant effect (taking $\Delta_p$ small compared to $\Delta_{1,2}$, for now at least), leading to a minimal energy configuration corresponding to the main saddle-point that has already been discussed in detail. However, this is not true when $x$ and $y$ approach one another, since the internal bulk-to-bulk propagator is singular, leading to a divergence in $\Gamma^{(p)}(x,y)$ scaling logarithmically with the distance, and a `tension' in the internal string diverging like the inverse of the distance. This large tension can balance the external operators, leading to an unstable equilibrium, our extra saddle-point of the energy functional.

The separation of the points $x$ and $y$ at this saddle-point is small, roughly the Compton wavelength of the external operators, so in particular much smaller than the AdS scale. To a first approximation, we can therefore ignore this internal structure, and treat the external particles as if they were joined at a tetravalent vertex at $x\approx y$, effectively giving a contact diagram. Minimising $S^{(1)}(x)+S^{(2)}(y)$ with $y=x$ finds the approximate location of the saddle-point, and gives the factor that scales exponentially with $\Delta_{1,2}$.

To calculate this explicitly, it is convenient to use a symmetric parameterisation of the operator insertion points, following \cite{1303.1111}. Putting the CFT on flat space, the conformal symmetry allows us to move all four insertion points to a plane parametrised by a complex coordinate $w$, to put the two insertions of $\phi_2$ at $w=\pm 1$, and to put the two insertions of $\phi_1$ at $w=\pm \rho$ for some complex $\rho$. This is simply related to the usual cross-ratio by  $x=\frac{4\rho}{(1+\rho)^2}$, or $\rho = \frac{x}{(1+\sqrt{1-x})^2}$. Use Poincar\'e coordinates for the bulk,
\begin{equation}
	ds^2 = \frac{dw d\bar{w}+dz^2+d\vec{x}^2}{z^2}
\end{equation}
with $\vec{x}$ denoting the $d-2$ directions transverse to the plane of the operator insertions, which will play no further r\^ole, and $z$ the bulk radial coordinate, with the boundary at $z=0$. Now, from symmetry, the vertices for the bulk geodesic network should lie on the axis $w=0$, $\vec{x}=0$, at some values of $z$ to be found. Computing the geodesic lengths (and subtracting the lengths corresponding to the factorised correlation function), the contact diagram with all geodesics meeting at the tetravalent vertex gives
\begin{equation}
	S^{(1)}(z)+S^{(2)}(z)=2 \Delta_1 \log \left(\frac{|\rho|^2+z^2}{2z|\rho|}\right)+2 \Delta_2 \log \left(\frac{z^2+1}{2z}\right),
\end{equation}
and to find stationary points we must solve $z^4+\frac{\Delta_1-\Delta_2}{\Delta_1+\Delta_2}(1-|\rho|^2)z^2-|\rho|^2=0$, which has a unique solution for positive $z$. At small $\rho$, the resulting contribution to the correlation function scales as $\rho^{2\min\{\Delta_1,\Delta_2\}}$, which corresponds to the exchange of the lightest double-trace operator.

To include the dependence on the internal dimension, it is a good approximation to treat the region around the vertices like flat space, with the scale of fluctuations of $x$ and $y$ around the saddle-point set by the Compton wavelength of the external particles, which is much less than the AdS scale. Integrating over $x$ and $y$ in this sub-AdS-scale region is equivalent to computing a flat space correlation function with the external operators taken to infinity in directions fixed by the previous calculation. The result is that the exchange diagram is corrected relative to the contact diagram by a flat-space tree level scattering amplitude, with external momenta fixed by an on-shell condition ($p^2+m^2=0$, so imaginary momentum in Euclidean signature, or Mandelstam variables $s,t,u$ all positive). This is reminiscent of the discussion in \cite{1007.2412}, in which a heavy particle could be `integrated out' by replacing its exchange diagrams with contact diagrams (including derivatives), normalised by flat-space amplitudes.

It remains to check that the details of this, perhaps by decomposing this extra saddle point contribution into conformal blocks, checking that it matches the double-trace contribution to the exchange diagram.

As a generalisation, we can allow the internal operator to also be heavy, $\Delta_p\gg 1$, taking a limit with all dimensions going to infinity at fixed ratios. As we increase the ratio of the internal to external dimensions, the dominant saddle-point shifts, with the vertices at $x$ and $y$ moving off the geodesics directly joining operator insertion points. It is straightforward to check that (in any dimension) the action \cref{networkAction} for the network at its minimum, with $\Gamma^{(p)}(x,y)$ just equal to $\Delta_p$ times the geodesic distance between $x$ and $y$, gives the global conformal block in the large $\Delta$ limit. Details of this calculation are given in \cref{largedim}. The other saddle-point, with $x$ and $y$ separated by a parametrically small distance, remains. This works until the internal dimension $\Delta_p$ becomes so large that the distance between $x$ and $y$ shrinks to zero. This corresponds to the point where the single-trace and double-trace contributions to the Witten diagram become the same size, and also to the point where the flat-space scattering amplitude describing the structure of the second saddle-point hits a pole, at $s=m_p^2$.

As $\Delta_p$ is increased beyond this critical value, the two saddle-points coalesce and disappear into the complex plane, leaving no real saddle-points for the path integral! We therefore are forced into including complex geodesics, as phenomenon observed previously, for example in \cite{hep-th/0306170}.

More generally, it would be interesting to understand the classification of all saddle-points, whether all or only some should be included on the path-integral contour, and how to sum them if there are infinitely many. For correlation functions, first steps in this direction were taken in \cite{1609.02165}.

\section{Discussion and outlook\label{disc}}

I conclude the paper by discussing how the new ideas introduced relate to other work, as well as various generalisations of the main approach, and outline some of the many future directions of study.

\subsection{Finite temperature\label{thermal}}

The discussion has focused on a small class of correlation functions in the vacuum state of the CFT, described by particles moving in pure AdS. But one of the virtues of the approach is that it immediately generalises to other spacetimes.

The simplest possible generalisation is to thermal AdS, dual to the CFT on a sphere $S^d$ at finite temperature, below the Hawking-Page transition. This is locally the same as pure AdS, simply with the time coordinate periodically identified, $t\sim t+\beta$. Without even adding external operators, we can compute quantities of interest, since the partition function encodes the spectrum of the theory. The contribution to the partition function of a free scalar has a nice interpretation using the worldline quantisation, which relates closely to the usual CFT intuition. It comes from the exponential of the sum of disconnected diagrams, which, without interactions, are worldlines describing closed loops in the spacetime. Since the spacetime has a nontrivial fundamental group, these worldlines are graded by their topology, namely the number $n$ of times that they wrap around the thermal circle. The $n=0$ sector gives a contribution proportional to the inverse temperature $\beta$, which, once exponentiated, simply shifts the vacuum energy; this should be absorbed in a renormalisation of $G_N$ and the cosmological constant, so we will drop it. In the $n=1$ sector, there is a unique geodesic, and fluctuations around this classical solution are computed by the partition function of the quantum-mechanics living on the worldline $Z_\text{WL}(\beta)$, computed using \cref{worldlineZ}. In each $n$ sector for $n\geq2$, there is still a unique geodesic, and this saddle-point contributes $Z_\text{WL}$, but evaluated at inverse temperature $n\beta$, modified by a symmetry factor of $\frac{1}{n}$ since fluctuations of the $n$ different loops are not physically distinct. Putting this together, we have
\begin{align*}
	Z(\beta)
	&=\exp\left(\sum_{n=1}^\infty \frac{1}{n} Z_\text{WL}(n\beta)\right) \\
	&= \exp\left( \mkern-4mu \sum_\text{WL states}\sum_{n=1}^\infty\frac{1}{n}e^{-nE}\right) \\
	&= \prod_\text{WL states} \frac{1}{1-e^{-\beta E}} \;,
\end{align*}
where we have written $Z_\text{WL}$ as a sum over states of energies $E$, and summed over $n$. The final result is the partition function of a Bose gas, with the terms in the product corresponding to different particle species, and expanding them %proddand
 as a geometric series, the powers in the expansion label the occupation number of particles of that species. In particular, $Z_{WL}$ is the partition-function of the single-particle states.

Finally, doing the worldline path integral gives $Z_{WL}(\beta)=\frac{e^{-\beta\Delta}}{(1-e^{-\beta})^d}$. To one loop on the worldline quantum mechanics, the denominator can be understood as coming from the partition function of $d$ simple harmonic oscillators\footnote{At this order, the integral over the modulus simply removes the contribution of the unphysical longitudinal oscillation $u$.}, which are the physical transverse fluctuations of the worldline. The ground state energies of the oscillators correct the exponent from $m$ to $m+\frac{d}{2}$, the first order expansion of $\Delta$. This matches the states of a single conformal multiplet in the CFT, with a primary of dimension $\Delta$ and descendants obtained by acting with derivatives. These derivatives, the momentum operators of the CFT in radial quantisation, are roughly the creation operators in the harmonic oscillators, and the annihilation operators correspond to special conformal generators. The Bose gas is what is expected of a generalised free field, with the various sectors of winding worldlines corresponding to multi-trace operators. To make a finer distinction between the multiple degenerate states, one could also turn on chemical potentials for rotations, identifying time periodically with a rotation, grading the quantum mechanics by its $SO(d)$ global symmetry.

This is simply a particle interpretation of the heat kernel method for computing partition functions, where the winding sectors correspond to the terms in the sum using the method of images to construct the heat kernel. As such, the same discussion applies in other circumstances such as higher genus handlebody spacetimes in $d=2$ \cite{0804.1773}.

If we turn on interactions of the particle with other fields, such as gravity, this will give corrections to the partition function that can be expanded diagrammatically using the methods of this paper. These will shift the energies of multi-particle states, corresponding to anomalous dimensions of the dual operators. This formalism therefore gives a natural and direct way to compute anomalous dimensions of operators, which will be developed in future work.

Moving on from the partition function, we can also compute other observables in thermal AdS, starting with finite-temperature one-point functions. These also can be decomposed in terms of conformal blocks, and in \cite{1706.00047}, these blocks were given an AdS interpretation. This gravity dual of a thermal block is very naturally stated in the language here, now with an internal operator $\phi$ quantised in the worldline formalism. The block is the contribution to the one-point function from the saddle-point where the internal particle worldline wraps the thermal circle once, truncating to single particle states, just as for the partition function. In the second quantised language, this means that we compute a Witten diagram, but instead of using the full thermal AdS propagator for $\phi$, constructed by the method of images, we truncate to only the pure AdS propagator. The usual Witten diagram includes contributions from multi-trace composites of $\phi$. It is rather natural in the worldline formalism that the block corresponds precisely to a single classical saddle-point, with all loop corrections. In this example, it is clear that the worldline loop corrections are vital, since restricting the vertex to lie in the centre of AdS does not give the correct answer for the block, unlike for the geodesic Witten diagrams.

Beyond thermal AdS, we can be more ambitious and apply these ideas to any background we choose, such as black holes, and the handlebody spacetimes already mentioned.

\subsection{Higher-point correlation functions and blocks}

In any given spacetime, we can also consider more complicated correlation functions, with more external operator insertions. Attempts to find bulk representations for 5-point and higher correlation functions, or even 2-point functions in thermal AdS, has proven tricky using the usual ingredients. The reason is relatively clear from the perspective of trying to find bulk objects that obey the same Casimir equations as conformal blocks: these follow fairly straightforwardly if the propagators in the bulk obey wave equations, but the usual propagators with the correct boundary conditions obey the wave equation with sources at coincident points. These sources must be avoided, somehow insulating the bulk points from ever touching, as is achieved by confining to geodesics, or by truncating the propagator in thermal AdS. This is hard to achieve in any covariant way for higher-point correlation functions.

The worldline formalism suggests a much more natural and general bulk dual of a conformal block, namely that the block is the all-loop perturbative contribution from a single saddle-point. This is most straightforward if we choose to quantise all relevant bulk fields on the worldline, so a bulk saddle-point corresponds to a network of geodesics joined at trivalent vertices (or higher-point if desired, though it is the trivalent vertex that is most relevant for conformal blocks), with the vertex locations chosen to minimise the worldline action (sum of masses times lengths). This was discussed briefly for four-point functions at the end of \cref{DT}, and in \cref{largedim}.

In \cite{1706.00047}, the classical action of such a configuration was shown to obey the same Casimir equations as global conformal blocks in a semiclassical limit taking all dimensions large, with fixed ratios\footnote{The proof was in the context of three dimensions, but I know of no obstruction in principle to generalising the argument to any dimension. Indeed, in the semiclassical limit the Casimir operators are insensitive to the dimension, assuming there are enough dimensions to properly represent the kinematics.}. This applies for any number of external operators, in any OPE channel, and in vacuum or a thermal background. Including quantum corrections to all orders on the worldline (in powers of $\Delta^{-1}$) from fluctuations of both the worldlines and the vertices at which they meet should promote the network to obey the full Casimir equations, not just the semiclassical versions. Following the intuition from the paragraph above, the edges will become full AdS propagators, obeying the wave equation, and the contact terms are not visible at any order in perturbation theory.

This requires some work to show rigorously, not least to give a definition of what the all-loop perturbation series means, since it is unlikely to be convergent; one might hope that the series is Borel summable, for example, in order to make this precise. A more straightforward avenue of enquiry is to develop the perturbation expansion with vertices in detail, in order to compute these objects to higher order, and compare to CFT expectations.

\subsection{OPE blocks and diffeomorphism invariant bulk observables \label{OPEblocks}}

A useful way of thinking about a geodesic Witten diagram, introduced in \cite{1604.03110}, is by splitting it up into components, writing it as a correlation function of a pair of `OPE blocks', which are nonlocal operators packaging the descendants of a particular primary $\phi_p$ appearing in the OPE of an operator $\phi$ with itself. This construction can be formally written as the integral of a free field in AdS dual to $\phi_p$, integrated over the bulk geodesic between the two operator insertion points. The correlation function of two of these objects then defines the conformal block, with the bulk representation, joining the fields by a bulk-to-bulk propagator, reproducing the geodesic Witten diagram recipe rather immediately. From the worldline point of view, it is natural to assign another set of words to the OPE block, namely that it is just the one-point function of the vertex operator! This seems to make sense only in perturbation theory, so perhaps it is better to identify the OPE block with the contribution to the vertex operator expectation value from a single saddle-point.

One of the most appealing aspects of the OPE block is that it represents a diffeomorphism invariant bulk observable with a relatively simple interpretation in the boundary CFT language. The connection with the worldline quantised particle is quite physical, with the particle (which in the classical limit has Compton wavelength much smaller than AdS scale, and hence is particularly sensitive to bulk locality) acting as a probe, and it also provides a natural framework in which to include quantum corrections (see \cite{1610.08952} for work in this direction).

In the present context, this is rather formal anyway, since strictly speaking, in Euclidean signature the OPE block has support on the entire boundary, rendering it largely useless since it cannot be inserted in any correlation function without contact terms. In Lorentzian signature, however, it is a much more natural object, and can be written with support only in a causal diamond. It seems that this connection should be revisited after better understanding the worldline quantisation in Lorentzian signature.

\subsection{Lorentzian kinematics}

This paper focussed attention on Euclidean signature, but of course many interesting questions involve real-time dynamics, so a pressing question is how to generalise the discussion to Lorentzian spacetime. In circumstances where a saddle-point is described by spacelike geodesics, there is no obvious obstruction to a direct generalisation, but this is not the most physical situation, and does not correspond to any time-ordered correlation function. A na\"ive generalisation for real-time correlation  functions of particles moving on timelike trajectories does not work straightforwardly, since a timelike geodesic never reaches the AdS boundary. This may be a sign of something physical, namely that a local operator insertion injects an infinite amount of energy into the CFT, so should correspond to an infinitely boosted particle moving along a null trajectory, which reaches the boundary. This divergent energy can be regulated by smoothing the insertion, perhaps by displacing the operator insertion by a small Euclidean time, but it is not immediately obvious how the worldline quantisation should then proceed.

These local (smeared) insertions of operators produce shockwave geometries, discussed in various contexts, including causality constraints, energy conditions, and chaos \cite{hep-th/0611122,0803.1467,1306.0622,1407.5597}, with various approaches compared recently in \cite{1709.03597}. The intuitive description of the bulk physics is suggestive that a worldline approach is natural, as are certain similarities in the discussions -- the `null energy' operator $\int du\, h_{uu}$ resembles the vertex operator one-point function, and double-trace contributions to the OPE are removed in both contexts. It would therefore be interesting to understand the connection.

This is also suggestive that the worldline perturbation theory may be valid in a different parametric regime from the one discussed in this paper. The small parameter controlling the classical limit of the worldline quantum mechanics here is the inverse dimension $\Delta^{-1}$ of the corresponding operator, but perhaps a different, kinematic, parameter could be used, and still have a good approximation with systematically computable corrections. For example, the exponentiation of the stress tensor observed in this paper is very similar to the observation in \cite{1504.01737}, except that in that work, the lightcone limit was relevant, rather than a large dimension limit.

\subsection{Virasoro blocks \label{VirDisc}}

I have already argued that the worldline quantised particle is the appropriate bulk definition of a Virasoro block in $d=2$. Here I expand slightly on that discussion, in particular pointing to various existing results and approaches in the literature.

The Virasoro blocks have been much studied in the `semiclassical' limit of $c\to\infty$ with $\frac{\Delta}{c}$ fixed, beginning with the work of Zamaloschikov \cite{zamolodchikov1987conformal} introducing the monodromy method for their computation, reviewed in \cite{1108.4417}. In this limit, the Virasoro blocks $V$ exponentiate, as
\begin{equation}
	V(\Delta,c;x)\sim e^{\frac{c}{6}f\left(\frac{\Delta}{c};x\right)},
\end{equation}
and furthermore, $f$ can be computed by solving a classical problem of heavy particles coupled to gravity. I do not know of any reference that proves this relationship in complete generality, but the works cited below provide examples in various special cases.

 One of the special features of gravity in $d=2$ is important in this context, namely that there is a minimal mass of order $c=\frac{3}{2G_N}$ below which black holes do not exist, and a lighter particle instead creates a conical defect along its worldline, similarly to a cosmic sting in four-dimensional flat spacetime. This black hole threshold corresponds to operator dimension $\Delta=\frac{c}{6}$, so we still have control of the worldline theory at finite $\frac{\Delta}{c}<\frac{1}{6}$ with the particle moving in a background created by its own backreaction with only a conical singularity, providing no obstruction to quantisation of the worldline theory\footnote{Above the threshold, the blocks are relevant for black hole physics, but I will not have anything to say about the blocks in this regime, except that the results above the threshold can be obtained by analytic continuation of those below.}. We can therefore talk sensibly about classical particles of finite mass coupled to gravity. It has been known for many years that the classical equations of motion of this system are equivalent to the Liouville equation with sources \cite{Deser:1983nh} (and see \cite{hep-th/0008253} for some discussion in the context of AdS/CFT). The same Liouville equation is also equivalent to the monodromy method calculation of the semiclassical blocks, with sources at operator insertions (indeed, one way to derive the semiclassical blocks is from the classical limit of Liouville CFT, governed by that equation).  In the slightly different context of R\'enyi entropies, this is discussed in \cite{1303.6955} from the CFT perspective, and \cite{1303.7221} from the bulk\footnote{This is in terms of a higher-genus replica surface, but is classically equivalent to talking about particles sourcing conical defects, after the Lewkowycz-Maldacena trick of taking a quotient by the replica symmetry \cite{1304.4926}. Quantum corrections to the higher genus blocks are, however, different from the Virasoro blocks with dimensions equal to twist operators \cite{1705.05855,Cho:2017fzo}.}.
 
This semiclassical limit is therefore equivalent to the tree-level diagrams discussed in this paper, which I have already argued are equivalent to the classical Einstein equations sourced by point particles. This problem is not exactly solvable in generality, so has been addressed in various limits, most notably by taking at most one of the particles (that is, two of the external operator insertions) to be heavy (with $\Delta/c$ of order one) \cite{1303.6955,1303.7221,1403.6829,1501.05315,1501.02260,1504.05943,1508.04987,1510.06685,1603.08440,1609.00801}, or by taking two particles to be heavy, but with a special value of the mass \cite{1609.02165}. A particularly interesting classical calculation in this context is \cite{1706.02668}, in which parametrically many operators are inserted, with fixed total energy, to create the dynamical background of a symmetric black hole formation.

A few discussions go beyond the strict semiclassical limit, with quantum corrections computed from the CFT using the Virasoro algebra \cite{1512.03052,1306.4682,1501.05315}, and in one instance, an expansion of the Virasoro algebra organised into diagrams very reminiscent of those in this paper \cite{Fitzpatrick:2015foa}, but there has been no general method to compute quantum corrections from the bulk proposed (except by using a Chern-Simons formulation, discussed in the next subsection). This problem was a key motivation for this paper, which provides a proposal for systematic computations of these quantum corrections.

I have not made any concerted effort to explicitly compute any of these quantum corrections in the current work, leaving this development for future study. Part of this development is to work out exactly how to best make use of the special properties of $d=2$. From the bulk side, three-dimensional gravity is simpler than higher dimensions, having no local bulk degrees of freedom, and the worldline quantum mechanics is one-loop exact, as mentioned at the end of \cref{WLpert}, so both of these can potentially be helpful. To reproduce and go beyond the results in the heavy-light limit, in which the dimension of one of the particles is taken to be of order $c$, requires taking a backreacted geometry, with a conical defect sourced by the heavy particle, as a starting-point for the perturbation theory. This requires a description of the quantum mechanics of a particle moving on this defect background.

\subsection{Connection to Chern-Simons formulation\label{CS}}

An alternative bulk dual of a Virasoro block was proposed in \cite{1612.06385}, building on work of Verlinde \cite{Verlinde:1989ua}, and applied in \cite{1702.06640} to compute self-energy corrections. This uses an $\mathfrak{sl}(2)$ Chern-Simons theory in the bulk,  which enjoys a close connection with 2+1 dimensional gravity with negative cosmological constant \cite{Witten:1988hc}. Wilson lines were first applied in the context of AdS/CFT to compute entanglement entropy in higher-spin theories \cite{1306.4338,1306.4347}. On the face of it, this construction looks very similar to the worldline formalism of this paper, just with the particle action written in gauge theory language as Wilson lines carrying an appropriate infinite-dimensional representation of the gauge group. This description has the advantage of a close relationship with the extended symmetry algebra of the CFT, which is far from manifest in my work.

However, it is not as straightforward to relate the two approaches as it might appear at first. One way of seeing this is to study the gauge symmetries of the Chern-Simons description and compare to gravity, written in terms of dreibein and spin connection. Half of the Chern-Simons gauge transformations are equivalent to the local Lorentz group, which is unproblematic. The other half are equivalent to diffeomorphisms, but \emph{only on-shell}, that is when evaluated on a flat connection, corresponding to a constant curvature geometry with vanishing torsion (the vanishing of the torsion of the connection is enforced by an equation of motion in the Chern-Simons formalism). This is not a problem for pure gravity, since one can directly quantise the phase space of classical solutions \cite{Witten:1988hc,1508.04079}, but becomes trickier when including matter, since then the matter sources the gauge connection so that it is not flat. Writing a gauge-invariant coupling in Chern-Simons language and na\"ively translating to gravity variables need not result in a diffeomorphism invariant action.

A hint of this comes even when considering global conformal blocks, when $G_N\to 0$ so gravity is non-dynamical and particles move on a fixed background. In the Chern-Simons language, conformal blocks can be described by the classical value of Wilson line networks with an appropriate flat connection \cite{1602.02962,1603.07317,1706.00047}. Since the connection is flat, this is independent of the path taken by the Wilson lines, in contrast to the alternative formulation in terms of worldline lengths, which depend on the details of the path taken. This is an indication that the notions of spacetime and locality in the two formulations are rather different, at least if taken at face value.

Finally, the vacuum Virasoro blocks do not give the full contribution to the correlation function in gravity, as discussed in \cref{DT} and elsewhere. Most simply, the order $\frac{1}{c}$ value of the four-point function $\langle\phi_1\phi_1\phi_2\phi_2\rangle$ comes from the graviton exchange Witten diagram, which includes double-trace exchanges as well as the stress tensor global conformal block. It is not immediately clear how or whether the Chern-Simons formulation accounts for these double-trace contributions.

It would be interesting to understand precisely how the two approaches are related, particularly as it may help to make a more direct connection between the bulk worldline description and the boundary Virasoro symmetry.

\subsection{Conformal bootstrap}

I conclude with a few comments and broad proposals for field theory motivated by the results of this paper, particularly in the context of the large-$N$ bootstrap. The most basic comment is that it suggests a new parameter regime to study, namely operators with dimensions that are large, but still parametrically below the gap. This is a natural context to study locality of the emergent bulk, because the physics in the gravitational dual is of particles with Compton wavelength parametrically smaller than AdS scale. Apart from some work in low dimension \cite{1611.10060}, this seems to be largely unexplored.

Some of the new results of the bulk calculations should be understandable in CFT language. In particular, it would be good to derive the fact that the global conformal block exponentiates from CFT consistency conditions, and appropriate sparseness assumptions, and to understand the consequences for OPE coefficients and anomalous dimensions of multi-trace operators.

\acknowledgments

It is a pleasure to thank Simon Caron-Huot and Alex Maloney for useful conversations and comments on a draft, and Alex Arvanitakis, Bartek Czech, Eliot Hijano, Nick Hunter-Jones, Lampros Lamprou, Sam McCandlish and Jamie Sully for helpful discussions.
I am supported by the Simons Foundation, through the It From Qubit collaboration.

\appendix

\section{Worldline QFT review}\label{WLreview}

In this appendix, I review the full derivation of the Polyakov gauge worldline path integral, \cref{WLPI}. My treatment will closely follow \cite{Polchinski:1998rq}, with the only generalisation being to allow the background to be curved. This amounts to little more than writing $g$ instead of $\eta$.

Following the ideas behind the quantisation of the string, but applied to particles, integrate over all paths on a target space with metric $g$, with `Polyakov' action:
\begin{equation}
	\mathcal{Z}=\int \frac{\piD{x}\piD{e}}{V_\text{diffs}} e^{-S[x,g]}; 
	\quad S_P[x,e,g]=\frac{1}{2}\int ds\, e(s) \left(e(s)^{-2}g_{ab}(x(s))\dot{x}^a(s)\dot{x}^b(s) + m^2 \right)
\end{equation}
The integral is over all functions $x$ from the worldline (which can either be an interval or $S^1$ depending on the context, the classification of one-manifolds being even simpler than for surfaces) to the target spacetime. This overcounts the paths as there is diffeomorphism invariance on the worldline, with any reparameterisation describing the same path in target space, so in the measure we formally divide out the volume of that gauge group. 

Now we want to gauge-fix this this action. First, an infintesimal diffeomorphism is represented by a vector field $\xi(s)\partial_s$, and acts on the fields as
\begin{equation}\label{WorldlineDiffs}
	\delta_\xi x =-\xi \dot{x}, \quad \delta_\xi e = -\dot{e}\xi-e \dot{\xi},
\end{equation}
with $x$ being a worldline scalar, and $e$ being a density. In particular, note that $\delta_\xi e$ is a derivative, so the integral of $e$ is invariant under diffeomorphisms that act trivially on the endpoints.

\subsection{Faddeev-Popov and BRST quantisation}
Now, recall the general Faddeev-Popov and BRST procedure, following \cite{Polchinski:1998rq}.

We have fields $\phi_i$ ($i$ here labels both fields ($x,e$) and coordinate $s$), an algebra of gauge transformations $[\delta_\alpha,\delta_\beta]=f^\gamma_{\phantom{\gamma}\alpha\beta}\delta_\gamma$, and gauge-fixing conditions $F^A(\phi)=0$. Define the Faddeev-Popov determinant by
\begin{equation}
	\Delta_{FP}(\phi)^{-1}=\int \piD{\zeta^\alpha} \prod_A \delta\left(F^A(\phi^\zeta)\right)
\end{equation}
where $\phi^\zeta=\zeta^\alpha\delta_\alpha\phi$ is the gauge transformation of the fields. This is gauge invariant assuming gauge invariance of the measure $\piD\zeta$. If there is a unique zero of the delta function, i.e.~the gauge orbits intersect the gauge-fixed slice in one location, $\Delta_{FP}$ is the determinant of the Jacobian $\delta_\alpha F^A$ there (in particular, the number of gauge parameters equals the number of gauge fixing conditions so the Jacobian is a `square matrix').

With this definition, we can insert unity into the path integral, with action $S(\phi)$, giving
\begin{align*}
	\int\frac{\piD \phi_i}{V_\text{gauge}}e^{-S(\phi)}&=\int\frac{\piD \phi_i \piD\zeta^\alpha}{V_\text{gauge}}e^{-S(\phi)}\Delta_{FP}(\phi)\prod_A \delta\left(F^A(\phi^\zeta)\right)\\
	&=\int\frac{\piD \phi_i^\zeta \piD\zeta^\alpha}{V_\text{gauge}}e^{-S(\phi^\zeta)}\Delta_{FP}(\phi^\zeta)\prod_A \delta\left(F^A(\phi^\zeta)\right)\\
	&=\int \piD \phi_i \,e^{-S(\phi)}\Delta_{FP}(\phi)\prod_A \delta\left(F^A(\phi)\right)
\end{align*}
where the second line uses gauge invariance of the measure, action and $\Delta_{FP}$ (as well as any other insertions into the path integral), and the third line changes variables and then integrates out the $\zeta$, removing the volume of the gauge group.

This gauge-fixing procedure has left us with a measure, which is the Fadeev-Popov determinant evaluated on a gauge-fixed $\phi$. Writing the determinant as a Grassmannian path integral over $b,c$, and representing the delta-function as a path integral over $B$, we finally have
\begin{align}
	\int & \piD{\phi_i}\piD{B_A}\piD{b_A}\piD{c^\alpha} \exp\left[-S(\phi)-S_{GF}(B,\phi)-S_{\text{ghost}}(b,c,\phi)\right]\\
	& S_{GF}(B,\phi)=-i B_A F^A(\phi),\quad S_{\text{ghost}}(b,c,\phi)=b_A c^\alpha \delta_\alpha F^A(\phi)
\end{align}
The gauge invariance leaves its mark as the BRST symmetry, acting as
\begin{align}
	[Q_B,\phi_i] =- c^\alpha \delta_\alpha \phi_i,&\quad
	[Q_B, B_A]=0,\nonumber \\
	\{Q_B, b_A\} = -i B_A,&\quad
	\{Q_B, c^\alpha\} = \frac{1}{2} f^\alpha_{\phantom{\alpha}\beta\gamma}c^\beta c^\gamma
\end{align}
which is nilpotent (by the Jacobi identity, in the case of $c$).
Varying the gauge-fixing condition should leave matrix elements of physical states invariant, but this is equivalent to adding a BRST exact term to the Hamiltonian. The physical states are therefore BRST closed (the physical Hilbert space is the cohomology of $Q_B$). It is often most convenient to integrate out $B$ by solving the constraints, replacing it in the BRST variation of $b$ by the equations of motion for the constrained fields.

\subsection{Polyakov gauge} \label{polyakov}
For the worldline action, the gauge fixing proceeds as above, with the gauge algebra coming from the one-dimensional diffs. The algebra is generated by functions on the interval or circle. The Polyakov gauge fixes the worldline parameter to be proportional to the proper length as measured with the auxiliary metric $e$, which means that we are setting $e$ to be a constant. I'll choose to set $e=m^{-1}$ (a slightly non-standard convention that keeps the units intact), so the gauge-fixing function is  $F^s[\phi]=m^{-1}-e(s)$. The ghost action is then
\begin{equation}
	S_\text{ghost} = \int ds \; b(s) \frac{d}{ds}(c(s)e(s)) =-\int ds\; e \dot{b} c
\end{equation}
which is the action for a simple two-level quantum system with vanishing Hamiltonian. In this gauge, the ghosts play essentially no r\^ole (completely independent of the theory on the worldline). This demonstrates that we can safely ignore the Fadeev-Popov ghosts in our calculations.
The BRST charge, computed by the N\"other procedure, is
\begin{equation}
	Q_B = \frac{1}{2} c (p^2+m^2),
\end{equation}
which imposes the mass-shell condition on physical states.

It only remains to observe that the gauge-fixing leaves a modulus $T=\int ds\,e(s)$ which must be integrated over, as discussed in the main text.

\section{The AdS heat kernel from worldline quantum mechanics\label{HKQM}}

In this appendix, I collect some details of the AdS worldline theory, specifically for calculations on a finite interval as relevant for computing the heat kernel in AdS by perturbative quantum mechanics. This includes summing the Feynman diagrams to find the heat kernel to two loops, \cref{twoLoopWL}.

\begin{table}
	\centering
	\begin{tabular}{m{2.7cm} m{12cm} }
		\begin{tikzpicture}
			\draw (0,0) -- (1,0) node [label=above:{$i$}]{} --(2,0);
			\path (0,-.6) -- (0,.6);
		\end{tikzpicture} & $ \lambda\, G_1(s_1,s_2)=\lambda\left[\frac{\sinh(L-s_1)\sinh s_2}{\sinh L}\Theta(s_1-s_2)+\frac{\sinh(L-s_2)\sinh s_1}{\sinh L}\Theta(s_2-s_1)\right] $ \\
		\begin{tikzpicture}
			\draw[dashed] (0,0) -- (2,0);
			\path (0,-.6) -- (0,.6);
		\end{tikzpicture} & $\lambda \,G_0(s_1,s_2)= \lambda\left[\frac{(L-s_1) s_2}{L}\Theta(s_1-s_2)+ \frac{(L-s_2) s_1}{L}\Theta(s_2-s_1)\right] $ \\
		\begin{tikzpicture}
			\draw [dashed]  (0,0) -- (-1,0) node [label=above:{$1$}]{};
			\draw (0,0) -- (.5, {sqrt(3)/2}) node [label=below:{$i$}]{};
			\draw (0,0) -- (.5, {-sqrt(3)/2}) node [label=above:{$j$}]{};
		\end{tikzpicture} & $-\frac{2}{\lambda}\delta_{ij} \partial_{s_1}$ \\
		\begin{tikzpicture}
			\draw [dashed]  (-.8,-.8) node [label=above:{$1$}]{} -- (0,0) -- (-.8,.8) node [label=below:{$2$}]{};
			\draw (0,0) -- (.8, .8) node [label=below:{$i$}]{};
			\draw (0,0) -- (.8, -.8) node [label=above:{$j$}]{};
		\end{tikzpicture} & $-\frac{2}{\lambda}\delta_{ij} \partial_{s_1}\partial_{s_2}$ \\
		\begin{tikzpicture}
			\draw (-.8,-.8) node [label=above:{$l$}]{} -- (0,0) -- (-.8,.8) node [label=below:{$i$}]{};
			\draw (0,0) -- (.8, .8) node [label=below:{$j$}]{};
			\draw (0,0) -- (.8, -.8) node [label=above:{$k$}]{};
		\end{tikzpicture} & $\frac{1}{\lambda}\left(\delta_{ij}\delta_{kl} \partial_{i}\partial_{l}+\text{11 permutations}\right)$ 
	\end{tabular}
	\caption{\label{FeynmanRules3}The Feynman rules for the AdS worldline quantum mechanics with boundary conditions at finite time, $q(0)=q(L)=u(0)=u(L)=0$. The $q$ propagators (indices $i,j,\ldots=1,2,\ldots,d$) are denoted with solid lines, and $u$ propagator with a dashed line. There are additional vertices similar to the final one for any even number of $q$ propagators.}
\end{table}

The Feynman rules are summarised in \cref{FeynmanRules3}. The vertices are the same as for the theory on the infinite line, but the propagators are different, solving the harmonic oscillator equation with a source, and boundary conditions at finite time: 

\begin{equation}
	(\omega^2-\partial_1^2)G_\omega(s_1,s_2)=\delta(s_1-s_2),\quad G_\omega(0,s_2)=G_\omega(L,s_2)=0
\end{equation} 

With these boundary conditions, the sum of vacuum diagrams computes the propagator of the AdS sigma model between points separated by proper distance $L$, the matrix element of $e^{\frac{L \lambda}{2}\nabla^2}$. This is the AdS heat kernel evaluated at distance $L$ and time $\frac{L\lambda}{2}$; the perturbative expansion in $\lambda$ is therefore a short-time expansion of the heat kernel.

At two loops, there are four vacuum diagrams, as well a a contribution from a counterterm required by the path integral regulator \cite{hep-th/9504097}. After drawing these diagrams, we can simply substitute the Feynman rules and compute the relevant integrals. The prefactors include symmetry factors, and factors of $d$ from counting the different $q$'s that can run in the loops.

Two of the diagrams are divergent, giving delta-functions evaluated at zero time difference, but these infinite contributions cancel. Going to higher loops, there are ambiguities from products of multiple distributions, that must be resolved with more careful inspection of the regulator \cite{hep-th/9504097}. Here, the diagrams have a minor ambiguity coming from evaluating the step function at 0, but this is resolved with the intuitive assignment $\Theta(0)=\frac{1}{2}$.

\begin{align*}
	\begin{tikzpicture}[baseline={([yshift=-.5ex]current bounding box.center)}]
			\node[draw,circle,minimum size=15](A) at (0,0){};
			\node[draw,circle,minimum size=15](B) at (1,0){};
			\draw[densely dashed] (A) --(B);
		\end{tikzpicture}
		&= \frac{d^2}{8} \int_0^L ds_1 ds_2 \; G_1(s_1,s_1)G_1(s_2,s_2) 4 \partial_1\partial_2 G_0(s_1,s_2) \\
		&= \frac{d^2}{16}\left(\frac{L}{\sinh^2 L}+\frac{1}{\tanh L}-\frac{2}{L}\right)  \displaybreak[1]\\
	\begin{tikzpicture}[baseline={([yshift=-.5ex]current bounding box.center)}]
			\draw (0,0) circle(.3);
			\draw[densely dashed] (-.3,0) -- (.3,0);
		\end{tikzpicture}
		&= \frac{d}{4} \int_0^L ds_1 ds_2 \; G_1(s_1,s_2)G_1(s_1,s_2) 4 \partial_1\partial_2 G_0(s_1,s_2) \\
		&= \frac{d}{8}\left(\frac{L}{\sinh^2 L}-\frac{5}{\tanh L}+2L+\frac{4}{L}\right) \displaybreak[1]\\
	\begin{tikzpicture}[baseline={([yshift=-.5ex]current bounding box.center)}]
			\draw[densely dashed] (0,0) to[out=45,in=90, looseness=1] (.8,0) to [out=-90,in=-45, looseness=1.5] (0,0);
			\draw (0,0) to[out=135,in=90, looseness=1] (-.8,0) to [out=-90,in=-135, looseness=1.5] (0,0);
		\end{tikzpicture}
		&= \frac{d}{4} \int_0^L ds \; G_1(s,s) (-2) \partial_1\partial_2 G_0(s,s) \\
		&= \frac{d}{4}(1-L\delta(0))\left(\frac{1}{\tanh L}-\frac{1}{L}\right) \displaybreak[1]\\
	\begin{tikzpicture}[baseline={([yshift=-.5ex]current bounding box.center)}]
			\draw (0,0) to[out=45,in=90, looseness=1] (.8,0) to [out=-90,in=-45, looseness=1.5] (0,0);
			\draw (0,0) to[out=135,in=90, looseness=1] (-.8,0) to [out=-90,in=-135, looseness=1.5] (0,0);
		\end{tikzpicture}
		&= \frac{1}{8} \int_0^L ds \; \left[4d G_1(s,s)\partial_1\partial_2 G_1(s,s) +4(d^2+d)\left(\partial_1 G_1(s,s)\right)^2\right] \\
		&= \frac{d}{4}L\delta(0)\left(\frac{1}{\tanh L}-\frac{1}{L}\right)+\frac{d}{16}\left(-\frac{L}{\sinh^2 L}+\frac{1}{\tanh L}-2L\right)\\
		&\qquad +\frac{d(d+1)}{16}\left(-\frac{L}{\sinh^2 L}+\frac{1}{\tanh L}\right)		
\end{align*}

We can now add all these up, but first we must account for the quantum counterterm in the action, $S_q=-\frac{\lambda}{8}\int_0^L ds(R+\Gamma\Gamma)$. Evaluating this on AdS in the coordinates used (since it is noncovariant), the contraction of the Christoffel symbols vanishes, and we have constant curvature $R=-d(d+1)$, resulting in the contribution $\frac{\lambda}{8} d(d+1)L$ to be added to the diagrams. Summing the diagrams along with this extra piece, the final result is
\begin{align*}
	W_L^{(2)}&=
	-\left(\begin{tikzpicture}[baseline={([yshift=-.5ex]current bounding box.center)}]
			\node[draw,circle,minimum size=15](A) at (0,0){};
			\node[draw,circle,minimum size=15](B) at (1,0){};
			\draw[densely dashed] (A) --(B);
		\end{tikzpicture}+
	\begin{tikzpicture}[baseline={([yshift=-.5ex]current bounding box.center)}]
			\draw (0,0) circle(.3);
			\draw[densely dashed] (-.3,0) -- (.3,0);
		\end{tikzpicture}
		+ \begin{tikzpicture}[baseline={([yshift=-.5ex]current bounding box.center)}]
			\draw (0,0) to[out=45,in=90, looseness=1] (.8,0) to [out=-90,in=-45, looseness=1.5] (0,0);
			\draw (0,0) to[out=135,in=90, looseness=1] (-.8,0) to [out=-90,in=-135, looseness=1.5] (0,0);
		\end{tikzpicture}
		+ \begin{tikzpicture}[baseline={([yshift=-.5ex]current bounding box.center)}]
			\draw[densely dashed] (0,0) to[out=45,in=90, looseness=1] (.8,0) to [out=-90,in=-45, looseness=1.5] (0,0);
			\draw (0,0) to[out=135,in=90, looseness=1] (-.8,0) to [out=-90,in=-135, looseness=1.5] (0,0);
		\end{tikzpicture}\right) + S_q
		 \nonumber\\
	 &=\frac{d(d-2)}{8}\left(\frac{1}{L}-\frac{1}{\tanh L}\right)+\frac{d^2}{8}L
\end{align*}
as claimed in \cref{twoLoopWL}.

\section{CFT and AdS in the large mass limit\label{largedim}}

In this appendix, I collect various useful results about conformal blocks, and AdS propagators and heat kernels, in a limit of large operator dimension or mass.

A convenient way to write Euclidean $AdS_{d+1}$ for this section is in polar coordinates, using the geodesic distance as a polar parameter,
\begin{equation}
	ds^2 = dx^2+\sinh^2 x \, d\Omega_d^2
\end{equation}
with $d\Omega_d^2$ the metric on the unit $d$-sphere. The Laplacian of a spherically symmetric field can then be written simply as
\begin{equation}
	\nabla^2 f(x) = \frac{1}{\sinh^d x}\frac{d}{dx}\left(\sinh^d x f'(x)\right)
\end{equation}

\subsection{AdS heat kernel}

Let us first look at the heat kernel, which is the solution to the heat equation with a delta-function initial condition:
\begin{equation}
	\frac{\partial K(x,t)}{\partial t} = \nabla^2 K(x,t),\quad K(x,t=0) = \delta^{(d+1)}(x)
\end{equation}
In hyperbolic space, putting the initial point at the origin of our coordinates, by symmetry it depends only on $x$ and $t$. The heat kernel in hyperbolic space admits closed form expressions, see for example \cite{grigor1998heat}, in terms of an integral for odd $d$. But the perturbative expansion of the weakly-coupled worldline theory corresponds to finding the short-time expansion of the heat kernel (an asymptotic series), which is straightforward to find simply by making an appropriate ansatz for $K$. The classical result corresponds to the flat-space heat kernel, with corrections in powers of $t$, so we make an ansatz
\begin{equation}
	K(x,t) = \frac{1}{(4\pi t)^\frac{d+1}{2}}e^{-\frac{x^2}{4t}-K_1(x)-K_2(x) t - K_3(x) t^2+\cdots}
\end{equation}
and expand in powers of $t$. At each order, imposing the heat equation results in a linear first-order equation in $x$, which we solve order by order, with the constant of integration chosen to avoid a singularity at $x = 0$.

To second order, the result is
\begin{equation}
	K(x,t) = \frac{1}{(4\pi t)^\frac{d+1}{2}}\left(\frac{x}{\sinh x}\right)^{d/2}e^{-\frac{x^2}{4t}} \left(1 + \left[\frac{d(d-2)}{4}\left(\frac{1}{x\tanh x}-\frac{1}{x^2}\right)-\frac{d^2}{4}\right]t+O(t^2)\right),
\end{equation}
matching the result computed from the quantum mechanics weak coupling expansion.

In three dimensions, the two-loop result is exact (with the contribution at two loops just shifting the ground state energy):
\begin{equation}
	K_{d=2}(x,t) = \frac{1}{(4\pi t)^\frac{3}{2}}\frac{x}{\sinh x} e^{-\frac{x^2}{4t}-t}
\end{equation}

\subsection{AdS propagator}

The Euclidean $AdS_{d+1}$ propagator for a scalar, with mass $m^2=\Delta(\Delta-d)$, is given by a hypergeometric function \cite{DHoker:2002nbb}
\begin{equation}\label{AdSprop}
	G_\Delta(x) = \frac{2^{-\Delta}\Gamma(\Delta)}{(2\Delta-d)\pi^{d/2}\Gamma\left(\Delta-\frac{d}{2}\right)} \left(\sech x
	\right)^\Delta {}_2F_1\left(\frac{\Delta}{2},\frac{\Delta+1}{2};\Delta+1-\frac{d}{2};\sech^2 x\right).
\end{equation}
where $x$ is the geodesic distance.

We now wish to expand this in the limit of large dimension $\Delta$. To do this, we can make an exponential ansatz for the equation of motion, $G_\Delta(x) = e^{\Delta G_0(x)+ G_1(x) + \Delta^{-1} G_2(x)+\cdots}$, and solve order by order, as for the heat kernel. As an illustration of an alternative technique (and to make it easier to check the normalisation), I find it with another method, using the hypergeometric function.
 
To do this, I'll use the integral representation of the hypergeometric function%
%(see \href{http://functions.wolfram.com/HypergeometricFunctions/Hypergeometric2F1/07/01/01/}{linky})
\begin{equation}
	_2F_1(a,b;c;z)=\frac{\Gamma(c)}{\Gamma(b)\Gamma(c-b)} \int_0^\infty t^{c-b-1}(1+t)^{a-c} (t+1-z)^{-a} \, dt\, ,
\end{equation}
which can be computed at large $a,b,c$ by saddle point.

Putting in the apropraiate values for $a,b,c$, we can write the integrand as
\begin{equation}
	f(t)\exp\left(-\frac{\Delta}{2}g(t)\right),\quad \text{with}\quad f(t)=t^{-\frac{1+d}{2}}(1+t)^{d/2-1},\quad g(t)=\log\left(\frac{(t+1)(t+\tanh^2x)}{t}\right)
\end{equation}
The function $g(t)$ has a single minimum at $t=t_0=\tanh x$, so we can do the saddle-point approximation around there. Going to two-loop order requires expanding $g$ to fourth order and $f$ to quadratic order around the saddle-point, and doing the Gaussian integrals with powers of $t-t_0$. For the integral, we get the expression
\begin{equation*}
	(1+\tanh(x))^{-\Delta}\sqrt{2\pi}(1+\coth x)^\frac{d}{2} \left[1+\left(\frac{d(d-2)}{8} \coth x-\frac{2d-1}{4}\right)\frac{1}{\Delta}+O(\Delta^{-2})\right]
\end{equation*}

Now we simply need to add in the $\Gamma$ function factors for the hypergeometric integral identity, as well as the prefactor in \cref{AdSprop}, and expand the $\Gamma$ functions using Stirling's approximation. One the dust settles, the result is
\begin{equation}
	G_\Delta(x) = \frac{(2\Delta)^{d/2-1}}{(2\pi)^{d/2}}\frac{e^{-\Delta\sigma}}{(1-e^{-2\sigma})^{d/2}} \left(1-\frac{d(d-2)}{8\Delta}\left(2-\frac{1}{\tanh x}\right)+O(\Delta)^{-2}\right)
\end{equation}
Alternatively, writing this in terms of the mass $m$, using $\Delta=m+\frac{d}{2}+\frac{d^2}{8m}+\cdots$, we get
\begin{equation}
	G_\Delta(x) = \frac{e^{-m L}}{2m}\left(\frac{m}{2\pi \sinh L}\right)^{d/2}\left(1+\frac{d(d-2)}{8m\tanh L}-\frac{d^2 L}{8m}+\cdots \right).
\end{equation}

%
%Penedones lecture notes \cite{Penedones:2016voo}: the bulk to bulk propagator is given in (73) as
%\begin{equation}
%	 G_\Delta = \frac{\Gamma(\Delta)}{2\pi^{d/2}\Gamma\left(\Delta-\frac{d}{2}+1\right)} \zeta^{-\Delta} {}_2F_1\left(\Delta,\Delta-\frac{d}{2}+\frac{1}{2};2\Delta-d+1;-\frac{4}{\zeta}\right)
%\end{equation}
%where $\zeta$ is the `chordal distance', which is $\zeta=2(\cosh x-1)$. Use the magic quadratic identity (15.8.13) \href{http://dlmf.nist.gov/15.8}{at this link} to get exactly the same result.

\subsection{Conformal blocks}

Moving away from AdS, I now give some results on conformal blocks in the large dimension limit.

For simplicity, focus on the case with external scalar operators, with dimensions equal in pairs, and define the blocks $\mathcal{F}$ such that the four-point function decomposes as
\begin{equation}
	\langle\phi_1(x_1)\phi_1(x_2)\phi_2(x_3)\phi_2(x_4)\rangle = \frac{1}{|x_1-x_2|^{2\Delta_1}} \frac{1}{|x_3-x_4|^{2\Delta_2}}\sum_p C_{11p}C_{22p} \mathcal{F}_p (x)
\end{equation}

Again, there are several ways to evaluate the blocks at large dimension, for example the saddle-point of an integral representation, but here I will use the Casimir equations satisfied by the blocks \cite{hep-th/0309180}. It's most convenient to use the radial $\rho,\bar{\rho}$ coordinates \cite{1303.1111}. Then, similarly to the heat kernel calculation, simply make an exponential ansatz for the Casimir differential equation and solve order by order in $\Delta$. The solutions at each order are not unique, with arbitrary functions of $\frac{\bar{\rho}}{\rho}$, but these can be fixed by the small-$\rho$ expansion, and the spin of the internal operator. For example, if the internal operator is a scalar, the result is
\begin{align*}
	\mathcal{F}_p(\rho,\bar{\rho})=\left(16\rho\bar{\rho}\right)^\frac{\Delta_p}{2} & \frac{(1-\rho\bar{\rho})^{1-\frac{d}{2}}}{\sqrt{(1-\rho^2)(1-\bar{\rho}^2)}}\\
	&\times\Big[1-\frac{1}{4\Delta_p}\left(\frac{2}{1-\rho^2}+\frac{2}{1-\bar{\rho}^2}-\frac{(d-2)(d-4)}{1-\rho  \bar{\rho}}+d^2-6d+4\right)+O(\Delta_p^{-2})\Big].
\end{align*}

\subsection{Geodesic networks and blocks}

Finally, I will give some details of the calculation showing that the scalar global conformal block in the large dimension limit, specifically the leading order of the result above, is obtained by minimising the worldline action for a geodesic network.

The action is roughly in \cref{networkAction}. Including the length of the exchanged geodesic, it is
\begin{equation*}
	S^{(1)}(z_1)+S^{(2)}(z_2)+S^{(p)}(z_1,z_2)=2 \Delta_1 \log \left(\frac{|\rho|^2+z_1^2}{2z_1|\rho|}\right)+2 \Delta_2 \log \left(\frac{z_2^2+1}{2z_2}\right)+\Delta_p \log\frac{z_2}{z_1}
\end{equation*}

Finding stationary points with respect to $z_1,z_2$, there is a unique solution
\begin{equation*}
	z_1 = \sqrt{\frac{2\Delta_1+\Delta_p}{2\Delta_1-\Delta_p}}\rho,\quad z_2 = \sqrt{\frac{2\Delta_2-\Delta_p}{2\Delta_2+\Delta_p}}
\end{equation*}
and, substituting this solution back in the action we find
\begin{equation}
	e^{-S} = \hat{C}(\Delta_1,\Delta_1,\Delta_p)\hat{C}(\Delta_2,\Delta_2,\Delta_p) \left(16\rho\bar{\rho}\right)^\frac{\Delta_p}{2},
\end{equation}
where $\hat{C}(\Delta_1,\Delta_2,\Delta_3)$ is a `semiclassical OPE coefficient' function, determined by computing a three-point function from a geodesic network. These coefficients were observed in the context of three dimensions \cite{1604.01774}, but the same values hold in general $d$. For general dimensions, the coefficients are given by
\begin{align}
        &\hat{C}(\Delta_i)=e^{\mathcal{P}(\Delta_i)},\text{ where} \\
        \mathcal{P}(\Delta_i)&=\frac{1}{2}\Delta_1\log\left[\frac{(\Delta_1+\Delta_2-\Delta_3)(\Delta_1+\Delta_3-\Delta_2)}{\Delta_2+\Delta_3-\Delta_1}\right] + (\text{2 permutations}) \nonumber \\
        &\quad\textstyle +\frac{1}{2}\left(\sum_i \Delta_i\right)(\log \sum_i\Delta_i-\log 4)-\sum_i \Delta_i\log\Delta_i \nonumber
\end{align}
and, for the special case of two identical dimensions as relevant here, we get
\begin{equation}
	\hat{C}(\Delta,\Delta,\Delta_p)=\left[\frac{(2\Delta-\Delta_p)(2\Delta+\Delta_p)}{4\Delta^2}\right]^\Delta \left[\frac{1}{4}\frac{2\Delta+\Delta_p}{2\Delta-\Delta_p}\right]^\frac{\Delta_p}{2}.
\end{equation}

The remaining piece of the action $(16\rho\bar{\rho})^\frac{\Delta_p}{2}$, containing all the dependence on the cross-ratio of the points, matches the tree-level piece of the global conformal block in the previous subsection. Note that the dimension $d$ drops out of the whole discussion.

Finally, this action was valid assuming $z_2>z_1$. If we take $\Delta_p$ to be too large, or the cross-ratio too close to one, the worldline of the exchanged particle shrinks to zero size. This regime is
\begin{equation}
	|\rho| > \sqrt{\frac{2\Delta_1-\Delta_p}{2\Delta_1+\Delta_p}\frac{2\Delta_2-\Delta_p}{2\Delta_2+\Delta_p}},
\end{equation}
in which the double-trace exchanges dominate over the single-trace.

\bibliographystyle{ssg}
\bibliography{worldlineBib2,worldlineBib}

\begingroup\raggedright\begin{thebibliography}{10}

\bibitem{hep-th/9711200}
J.~M. Maldacena, ``{The Large N limit of superconformal field theories and
  supergravity},'' {\em Int. J. Theor. Phys.} {\bf 38} (1999) 1113--1133,
  \href{http://xxx.lanl.gov/abs/hep-th/9711200}{{\tt hep-th/9711200}}. [Adv.
  Theor. Math. Phys.2,231(1998)].

\bibitem{hep-th/9802109}
S.~S. Gubser, I.~R. Klebanov, and A.~M. Polyakov, ``{Gauge theory correlators
  from noncritical string theory},'' {\em Phys. Lett.} {\bf B428} (1998)
  105--114, \href{http://xxx.lanl.gov/abs/hep-th/9802109}{{\tt
  hep-th/9802109}}.

\bibitem{hep-th/9802150}
E.~Witten, ``{Anti-de Sitter space and holography},'' {\em Adv. Theor. Math.
  Phys.} {\bf 2} (1998) 253--291,
  \href{http://xxx.lanl.gov/abs/hep-th/9802150}{{\tt hep-th/9802150}}.

\bibitem{Hawking:1976ra}
S.~W. Hawking, ``{Breakdown of Predictability in Gravitational Collapse},''
  {\em Phys. Rev.} {\bf D14} (1976) 2460--2473.

\bibitem{0907.0151}
I.~Heemskerk, J.~Penedones, J.~Polchinski, and J.~Sully, ``{Holography from
  Conformal Field Theory},'' {\em JHEP} {\bf 10} (2009) 079,
  \href{http://xxx.lanl.gov/abs/0907.0151}{{\tt 0907.0151}}.

\bibitem{1101.4163}
S.~El-Showk and K.~Papadodimas, ``{Emergent Spacetime and Holographic CFTs},''
  {\em JHEP} {\bf 10} (2012) 106, \href{http://xxx.lanl.gov/abs/1101.4163}{{\tt
  1101.4163}}.

\bibitem{1207.3123}
A.~Almheiri, D.~Marolf, J.~Polchinski, and J.~Sully, ``{Black Holes:
  Complementarity or Firewalls?},'' {\em JHEP} {\bf 02} (2013) 062,
  \href{http://xxx.lanl.gov/abs/1207.3123}{{\tt 1207.3123}}.

\bibitem{1304.6483}
A.~Almheiri, D.~Marolf, J.~Polchinski, D.~Stanford, and J.~Sully, ``{An
  Apologia for Firewalls},'' {\em JHEP} {\bf 09} (2013) 018,
  \href{http://xxx.lanl.gov/abs/1304.6483}{{\tt 1304.6483}}.

\bibitem{1612.03891}
O.~Aharony, L.~F. Alday, A.~Bissi, and E.~Perlmutter, ``{Loops in AdS from
  Conformal Field Theory},'' {\em JHEP} {\bf 07} (2017) 036,
  \href{http://xxx.lanl.gov/abs/1612.03891}{{\tt 1612.03891}}.

\bibitem{1612.06385}
A.~L. Fitzpatrick, J.~Kaplan, D.~Li, and J.~Wang, ``{Exact Virasoro Blocks from
  Wilson Lines and Background-Independent Operators},'' {\em JHEP} {\bf 07}
  (2017) 092, \href{http://xxx.lanl.gov/abs/1612.06385}{{\tt 1612.06385}}.

\bibitem{Feynman:1950ir}
R.~P. Feynman, ``{Mathematical formulation of the quantum theory of
  electromagnetic interaction},'' {\em Phys. Rev.} {\bf 80} (1950) 440--457.

\bibitem{Bern:1987tw}
Z.~Bern and D.~A. Kosower, ``{A New Approach to One Loop Calculations in Gauge
  Theories},'' {\em Phys. Rev.} {\bf D38} (1988) 1888.

\bibitem{hep-ph/9205205}
M.~J. Strassler, ``{Field theory without Feynman diagrams: One loop effective
  actions},'' {\em Nucl. Phys.} {\bf B385} (1992) 145--184,
  \href{http://xxx.lanl.gov/abs/hep-ph/9205205}{{\tt hep-ph/9205205}}.

\bibitem{AlvarezGaume:1983ig}
L.~Alvarez-Gaume and E.~Witten, ``{Gravitational Anomalies},'' {\em Nucl.
  Phys.} {\bf B234} (1984) 269.

\bibitem{hep-th/0101036}
C.~Schubert, ``{Perturbative quantum field theory in the string inspired
  formalism},'' {\em Phys. Rept.} {\bf 355} (2001) 73--234,
  \href{http://xxx.lanl.gov/abs/hep-th/0101036}{{\tt hep-th/0101036}}.

\bibitem{1508.00501}
E.~Hijano, P.~Kraus, E.~Perlmutter, and R.~Snively, ``{Witten Diagrams
  Revisited: The AdS Geometry of Conformal Blocks},'' {\em JHEP} {\bf 01}
  (2016) 146, \href{http://xxx.lanl.gov/abs/1508.00501}{{\tt 1508.00501}}.

\bibitem{Bastianelli:2006rx}
F.~Bastianelli and P.~van Nieuwenhuizen, {\em {Path integrals and anomalies in
  curved space}}.
\newblock Cambridge University Press, 2006.

\bibitem{Polyakov:1987ez}
A.~M. Polyakov, ``{Gauge Fields and Strings},'' {\em Contemp. Concepts Phys.}
  {\bf 3} (1987) 1--301.

\bibitem{hep-th/9112035}
F.~Bastianelli, ``{The Path integral for a particle in curved spaces and Weyl
  anomalies},'' {\em Nucl. Phys.} {\bf B376} (1992) 113--126,
  \href{http://xxx.lanl.gov/abs/hep-th/9112035}{{\tt hep-th/9112035}}.

\bibitem{hep-th/9504097}
J.~De~Boer, B.~Peeters, K.~Skenderis, and P.~Van~Nieuwenhuizen, ``{Loop
  calculations in quantum mechanical nonlinear sigma models},'' {\em Nucl.
  Phys.} {\bf B446} (1995) 211--222,
  \href{http://xxx.lanl.gov/abs/hep-th/9504097}{{\tt hep-th/9504097}}.

\bibitem{duistermaat1982variation}
J.~J. Duistermaat and G.~J. Heckman, ``On the variation in the cohomology of
  the symplectic form of the reduced phase space,'' {\em Inventiones
  mathematicae} {\bf 69} (1982), no.~2 259--268.

\bibitem{Picken:1988ev}
R.~F. Picken, ``{The Propagator for Quantum Mechanics on a Group Manifold From
  an Infinite Dimensional Analog of the Duistermaat-heckman Integration
  Formula},'' {\em J. Phys.} {\bf A22} (1989) 2285.

\bibitem{hep-th/9608068}
R.~J. Szabo, ``{Equivariant localization of path integrals},''
  \href{http://xxx.lanl.gov/abs/hep-th/9608068}{{\tt hep-th/9608068}}.

\bibitem{1703.04612}
D.~Stanford and E.~Witten, ``{Fermionic Localization of the Schwarzian
  Theory},'' {\em JHEP} {\bf 10} (2017) 008,
  \href{http://xxx.lanl.gov/abs/1703.04612}{{\tt 1703.04612}}.

\bibitem{grigor1998heat}
A.~Grigor'yan and M.~Noguchi, ``The heat kernel on hyperbolic space,'' {\em
  Bulletin of the London Mathematical Society} {\bf 30} (1998), no.~6 643--650.

\bibitem{Brink:1976sz}
L.~Brink, S.~Deser, B.~Zumino, P.~Di~Vecchia, and P.~S. Howe, ``{Local
  Supersymmetry for Spinning Particles},'' {\em Phys. Lett.} {\bf 64B} (1976)
  435. [Erratum: Phys. Lett.68B,488(1977)].

\bibitem{Brink:1976uf}
L.~Brink, P.~Di~Vecchia, and P.~S. Howe, ``{A Lagrangian Formulation of the
  Classical and Quantum Dynamics of Spinning Particles},'' {\em Nucl. Phys.}
  {\bf B118} (1977) 76--94.

\bibitem{1706.00047}
P.~Kraus, A.~Maloney, H.~Maxfield, G.~S. Ng, and J.-q. Wu, ``{Witten Diagrams
  for Torus Conformal Blocks},'' {\em JHEP} {\bf 09} (2017) 149,
  \href{http://xxx.lanl.gov/abs/1706.00047}{{\tt 1706.00047}}.

\bibitem{1604.03110}
B.~Czech, L.~Lamprou, S.~McCandlish, B.~Mosk, and J.~Sully, ``{A Stereoscopic
  Look into the Bulk},'' {\em JHEP} {\bf 07} (2016) 129,
  \href{http://xxx.lanl.gov/abs/1604.03110}{{\tt 1604.03110}}.

\bibitem{1403.6829}
A.~L. Fitzpatrick, J.~Kaplan, and M.~T. Walters, ``{Universality of
  Long-Distance AdS Physics from the CFT Bootstrap},'' {\em JHEP} {\bf 08}
  (2014) 145, \href{http://xxx.lanl.gov/abs/1403.6829}{{\tt 1403.6829}}.

\bibitem{1504.01737}
A.~L. Fitzpatrick, J.~Kaplan, M.~T. Walters, and J.~Wang, ``{Eikonalization of
  Conformal Blocks},'' {\em JHEP} {\bf 09} (2015) 019,
  \href{http://xxx.lanl.gov/abs/1504.01737}{{\tt 1504.01737}}.

\bibitem{hep-th/9902042}
E.~D'Hoker, D.~Z. Freedman, S.~D. Mathur, A.~Matusis, and L.~Rastelli,
  ``{Graviton and gauge boson propagators in AdS(d+1)},'' {\em Nucl. Phys.}
  {\bf B562} (1999) 330--352,
  \href{http://xxx.lanl.gov/abs/hep-th/9902042}{{\tt hep-th/9902042}}.

\bibitem{einstein1914erklarung}
A.~Einstein, ``Erkl{\"a}rung der Perihelbewegung des Merkur aus der allgemeinen
  Relativit{\"a}tstheorie,'' {\em Albert Einstein: Akademie-Vortr{\"a}ge:
  Sitzungsberichte der Preu{\ss}ischen Akademie der Wissenschaften 1914-1932}
  78--87.

\bibitem{Abbott:2016blz}
{\bf Virgo, LIGO Scientific} Collaboration, B.~P. Abbott {\em et.~al.},
  ``{Observation of Gravitational Waves from a Binary Black Hole Merger},''
  {\em Phys. Rev. Lett.} {\bf 116} (2016), no.~6 061102,
  \href{http://xxx.lanl.gov/abs/1602.03837}{{\tt 1602.03837}}.

\bibitem{hep-th/0409156}
W.~D. Goldberger and I.~Z. Rothstein, ``{An Effective field theory of gravity
  for extended objects},'' {\em Phys. Rev.} {\bf D73} (2006) 104029,
  \href{http://xxx.lanl.gov/abs/hep-th/0409156}{{\tt hep-th/0409156}}.

\bibitem{1303.6955}
T.~Hartman, ``{Entanglement Entropy at Large Central Charge},''
  \href{http://xxx.lanl.gov/abs/1303.6955}{{\tt 1303.6955}}.

\bibitem{1502.07742}
E.~Perlmutter, ``{Virasoro conformal blocks in closed form},'' {\em JHEP} {\bf
  08} (2015) 088, \href{http://xxx.lanl.gov/abs/1502.07742}{{\tt 1502.07742}}.

\bibitem{1703.00278}
S.~Caron-Huot, ``{Analyticity in Spin in Conformal Theories},'' {\em JHEP} {\bf
  09} (2017) 078, \href{http://xxx.lanl.gov/abs/1703.00278}{{\tt 1703.00278}}.

\bibitem{hep-th/9903196}
E.~D'Hoker, D.~Z. Freedman, S.~D. Mathur, A.~Matusis, and L.~Rastelli,
  ``{Graviton exchange and complete four point functions in the AdS / CFT
  correspondence},'' {\em Nucl. Phys.} {\bf B562} (1999) 353--394,
  \href{http://xxx.lanl.gov/abs/hep-th/9903196}{{\tt hep-th/9903196}}.

\bibitem{1303.1111}
M.~Hogervorst and S.~Rychkov, ``{Radial Coordinates for Conformal Blocks},''
  {\em Phys. Rev.} {\bf D87} (2013) 106004,
  \href{http://xxx.lanl.gov/abs/1303.1111}{{\tt 1303.1111}}.

\bibitem{1007.2412}
A.~L. Fitzpatrick, E.~Katz, D.~Poland, and D.~Simmons-Duffin, ``{Effective
  Conformal Theory and the Flat-Space Limit of AdS},'' {\em JHEP} {\bf 07}
  (2011) 023, \href{http://xxx.lanl.gov/abs/1007.2412}{{\tt 1007.2412}}.

\bibitem{hep-th/0306170}
L.~Fidkowski, V.~Hubeny, M.~Kleban, and S.~Shenker, ``{The Black hole
  singularity in AdS / CFT},'' {\em JHEP} {\bf 02} (2004) 014,
  \href{http://xxx.lanl.gov/abs/hep-th/0306170}{{\tt hep-th/0306170}}.

\bibitem{1609.02165}
A.~Maloney, H.~Maxfield, and G.~S. Ng, ``{A conformal block Farey tail},'' {\em
  JHEP} {\bf 06} (2017) 117, \href{http://xxx.lanl.gov/abs/1609.02165}{{\tt
  1609.02165}}.

\bibitem{0804.1773}
S.~Giombi, A.~Maloney, and X.~Yin, ``{One-loop Partition Functions of 3D
  Gravity},'' {\em JHEP} {\bf 08} (2008) 007,
  \href{http://xxx.lanl.gov/abs/0804.1773}{{\tt 0804.1773}}.

\bibitem{1610.08952}
M.~Guica, ``{Bulk fields from the boundary OPE},''
  \href{http://xxx.lanl.gov/abs/1610.08952}{{\tt 1610.08952}}.

\bibitem{hep-th/0611122}
L.~Cornalba, M.~S. Costa, J.~Penedones, and R.~Schiappa, ``{Eikonal
  Approximation in AdS/CFT: From Shock Waves to Four-Point Functions},'' {\em
  JHEP} {\bf 08} (2007) 019, \href{http://xxx.lanl.gov/abs/hep-th/0611122}{{\tt
  hep-th/0611122}}.

\bibitem{0803.1467}
D.~M. Hofman and J.~Maldacena, ``{Conformal collider physics: Energy and charge
  correlations},'' {\em JHEP} {\bf 05} (2008) 012,
  \href{http://xxx.lanl.gov/abs/0803.1467}{{\tt 0803.1467}}.

\bibitem{1306.0622}
S.~H. Shenker and D.~Stanford, ``{Black holes and the butterfly effect},'' {\em
  JHEP} {\bf 03} (2014) 067, \href{http://xxx.lanl.gov/abs/1306.0622}{{\tt
  1306.0622}}.

\bibitem{1407.5597}
X.~O. Camanho, J.~D. Edelstein, J.~Maldacena, and A.~Zhiboedov, ``{Causality
  Constraints on Corrections to the Graviton Three-Point Coupling},'' {\em
  JHEP} {\bf 02} (2016) 020, \href{http://xxx.lanl.gov/abs/1407.5597}{{\tt
  1407.5597}}.

\bibitem{1709.03597}
N.~Afkhami-Jeddi, T.~Hartman, S.~Kundu, and A.~Tajdini, ``{Shockwaves from the
  Operator Product Expansion},'' \href{http://xxx.lanl.gov/abs/1709.03597}{{\tt
  1709.03597}}.

\bibitem{zamolodchikov1987conformal}
A.~B. Zamolodchikov, ``Conformal symmetry in two-dimensional space: recursion
  representation of conformal block,'' {\em Theoretical and Mathematical
  Physics} {\bf 73} (1987), no.~1 1088--1093.

\bibitem{1108.4417}
D.~Harlow, J.~Maltz, and E.~Witten, ``{Analytic Continuation of Liouville
  Theory},'' {\em JHEP} {\bf 12} (2011) 071,
  \href{http://xxx.lanl.gov/abs/1108.4417}{{\tt 1108.4417}}.

\bibitem{Deser:1983nh}
S.~Deser and R.~Jackiw, ``{Three-Dimensional Cosmological Gravity: Dynamics of
  Constant Curvature},'' {\em Annals Phys.} {\bf 153} (1984) 405--416.

\bibitem{hep-th/0008253}
K.~Krasnov, ``{3-D gravity, point particles and Liouville theory},'' {\em
  Class. Quant. Grav.} {\bf 18} (2001) 1291--1304,
  \href{http://xxx.lanl.gov/abs/hep-th/0008253}{{\tt hep-th/0008253}}.

\bibitem{1303.7221}
T.~Faulkner, ``{The Entanglement Renyi Entropies of Disjoint Intervals in
  AdS/CFT},'' \href{http://xxx.lanl.gov/abs/1303.7221}{{\tt 1303.7221}}.

\bibitem{1304.4926}
A.~Lewkowycz and J.~Maldacena, ``{Generalized gravitational entropy},'' {\em
  JHEP} {\bf 08} (2013) 090, \href{http://xxx.lanl.gov/abs/1304.4926}{{\tt
  1304.4926}}.

\bibitem{1705.05855}
J.~Cardy, A.~Maloney, and H.~Maxfield, ``{A new handle on three-point
  coefficients: OPE asymptotics from genus two modular invariance},'' {\em
  JHEP} {\bf 10} (2017) 136, \href{http://xxx.lanl.gov/abs/1705.05855}{{\tt
  1705.05855}}.

\bibitem{Cho:2017fzo}
M.~Cho, S.~Collier, and X.~Yin, ``{Genus Two Modular Bootstrap},''
  \href{http://xxx.lanl.gov/abs/1705.05865}{{\tt 1705.05865}}.

\bibitem{1501.05315}
A.~L. Fitzpatrick, J.~Kaplan, and M.~T. Walters, ``{Virasoro Conformal Blocks
  and Thermality from Classical Background Fields},'' {\em JHEP} {\bf 11}
  (2015) 200, \href{http://xxx.lanl.gov/abs/1501.05315}{{\tt 1501.05315}}.

\bibitem{1501.02260}
E.~Hijano, P.~Kraus, and R.~Snively, ``{Worldline approach to semi-classical
  conformal blocks},'' {\em JHEP} {\bf 07} (2015) 131,
  \href{http://xxx.lanl.gov/abs/1501.02260}{{\tt 1501.02260}}.

\bibitem{1504.05943}
K.~B. Alkalaev and V.~A. Belavin, ``{Classical conformal blocks via AdS/CFT
  correspondence},'' {\em JHEP} {\bf 08} (2015) 049,
  \href{http://xxx.lanl.gov/abs/1504.05943}{{\tt 1504.05943}}.

\bibitem{1508.04987}
E.~Hijano, P.~Kraus, E.~Perlmutter, and R.~Snively, ``{Semiclassical Virasoro
  blocks from AdS$_{3}$ gravity},'' {\em JHEP} {\bf 12} (2015) 077,
  \href{http://xxx.lanl.gov/abs/1508.04987}{{\tt 1508.04987}}.

\bibitem{1510.06685}
K.~B. Alkalaev and V.~A. Belavin, ``{Monodromic vs geodesic computation of
  Virasoro classical conformal blocks},'' {\em Nucl. Phys.} {\bf B904} (2016)
  367--385, \href{http://xxx.lanl.gov/abs/1510.06685}{{\tt 1510.06685}}.

\bibitem{1603.08440}
K.~B. Alkalaev and V.~A. Belavin, ``{Holographic interpretation of 1-point
  toroidal block in the semiclassical limit},'' {\em JHEP} {\bf 06} (2016) 183,
  \href{http://xxx.lanl.gov/abs/1603.08440}{{\tt 1603.08440}}.

\bibitem{1609.00801}
B.~Chen, J.-q. Wu, and J.-j. Zhang, ``{Holographic Description of 2D Conformal
  Block in Semi-classical Limit},'' {\em JHEP} {\bf 10} (2016) 110,
  \href{http://xxx.lanl.gov/abs/1609.00801}{{\tt 1609.00801}}.

\bibitem{1706.02668}
T.~Anous, T.~Hartman, A.~Rovai, and J.~Sonner, ``{From Conformal Blocks to Path
  Integrals in the Vaidya Geometry},'' {\em JHEP} {\bf 09} (2017) 009,
  \href{http://xxx.lanl.gov/abs/1706.02668}{{\tt 1706.02668}}.

\bibitem{1512.03052}
A.~L. Fitzpatrick and J.~Kaplan, ``{Conformal Blocks Beyond the Semi-Classical
  Limit},'' {\em JHEP} {\bf 05} (2016) 075,
  \href{http://xxx.lanl.gov/abs/1512.03052}{{\tt 1512.03052}}.

\bibitem{1306.4682}
T.~Barrella, X.~Dong, S.~A. Hartnoll, and V.~L. Martin, ``{Holographic
  entanglement beyond classical gravity},'' {\em JHEP} {\bf 09} (2013) 109,
  \href{http://xxx.lanl.gov/abs/1306.4682}{{\tt 1306.4682}}.

\bibitem{Fitzpatrick:2015foa}
A.~L. Fitzpatrick, J.~Kaplan, M.~T. Walters, and J.~Wang, ``{Hawking from
  Catalan},'' {\em JHEP} {\bf 05} (2016) 069,
  \href{http://xxx.lanl.gov/abs/1510.00014}{{\tt 1510.00014}}.

\bibitem{Verlinde:1989ua}
H.~L. Verlinde, ``{Conformal Field Theory, 2-$D$ Quantum Gravity and
  Quantization of Teichmuller Space},'' {\em Nucl. Phys.} {\bf B337} (1990)
  652--680.

\bibitem{1702.06640}
M.~Besken, A.~Hegde, and P.~Kraus, ``{Anomalous dimensions from quantum Wilson
  lines},'' \href{http://xxx.lanl.gov/abs/1702.06640}{{\tt 1702.06640}}.

\bibitem{Witten:1988hc}
E.~Witten, ``{(2+1)-Dimensional Gravity as an Exactly Soluble System},'' {\em
  Nucl. Phys.} {\bf B311} (1988) 46.

\bibitem{1306.4338}
M.~Ammon, A.~Castro, and N.~Iqbal, ``{Wilson Lines and Entanglement Entropy in
  Higher Spin Gravity},'' {\em JHEP} {\bf 10} (2013) 110,
  \href{http://xxx.lanl.gov/abs/1306.4338}{{\tt 1306.4338}}.

\bibitem{1306.4347}
J.~de~Boer and J.~I. Jottar, ``{Entanglement Entropy and Higher Spin Holography
  in AdS$_3$},'' {\em JHEP} {\bf 04} (2014) 089,
  \href{http://xxx.lanl.gov/abs/1306.4347}{{\tt 1306.4347}}.

\bibitem{1508.04079}
A.~Maloney, ``{Geometric Microstates for the Three Dimensional Black Hole?},''
  \href{http://xxx.lanl.gov/abs/1508.04079}{{\tt 1508.04079}}.

\bibitem{1602.02962}
A.~Bhatta, P.~Raman, and N.~V. Suryanarayana, ``{Holographic Conformal Partial
  Waves as Gravitational Open Wilson Networks},'' {\em JHEP} {\bf 06} (2016)
  119, \href{http://xxx.lanl.gov/abs/1602.02962}{{\tt 1602.02962}}.

\bibitem{1603.07317}
M.~Besken, A.~Hegde, E.~Hijano, and P.~Kraus, ``{Holographic conformal blocks
  from interacting Wilson lines},'' {\em JHEP} {\bf 08} (2016) 099,
  \href{http://xxx.lanl.gov/abs/1603.07317}{{\tt 1603.07317}}.

\bibitem{1611.10060}
D.~Mazac, ``{Analytic bounds and emergence of AdS$_{2}$ physics from the
  conformal bootstrap},'' {\em JHEP} {\bf 04} (2017) 146,
  \href{http://xxx.lanl.gov/abs/1611.10060}{{\tt 1611.10060}}.

\bibitem{Polchinski:1998rq}
J.~Polchinski, {\em {String theory. Vol. 1: An introduction to the bosonic
  string}}.
\newblock Cambridge University Press, 2007.

\bibitem{DHoker:2002nbb}
E.~D'Hoker and D.~Z. Freedman, ``{Supersymmetric gauge theories and the AdS /
  CFT correspondence},'' in {\em {Strings, Branes and Extra Dimensions: TASI
  2001: Proceedings}}, pp.~3--158, 2002.
\newblock \href{http://xxx.lanl.gov/abs/hep-th/0201253}{{\tt hep-th/0201253}}.

\bibitem{hep-th/0309180}
F.~A. Dolan and H.~Osborn, ``{Conformal partial waves and the operator product
  expansion},'' {\em Nucl. Phys.} {\bf B678} (2004) 491--507,
  \href{http://xxx.lanl.gov/abs/hep-th/0309180}{{\tt hep-th/0309180}}.

\bibitem{1604.01774}
C.-M. Chang and Y.-H. Lin, ``{Bootstrap, universality and horizons},'' {\em
  JHEP} {\bf 10} (2016) 068, \href{http://xxx.lanl.gov/abs/1604.01774}{{\tt
  1604.01774}}.

\end{thebibliography}\endgroup

\end{document}